\documentclass[11pt,reqno]{amsart}
\usepackage{amsfonts}
\usepackage{mathrsfs}
\usepackage{graphicx} %We can use any other package if it is necessary
\usepackage{epsfig}
\usepackage{amsmath, amsthm}
\usepackage{amssymb}
\usepackage[all]{xy}
\usepackage{color}
\numberwithin{equation}{section}
\date{}

\theoremstyle{plain}

\numberwithin{equation}{section}

\newtheorem{thm}{Theorem}[section]
\newtheorem{prop}[thm]{Proposition}
\newtheorem{lem}[thm]{Lemma}
\newtheorem{cor}[thm]{Corollary}

\theoremstyle{definition}

\newtheorem{rem}[thm]{Remark}
\newtheorem{defn}[thm]{Definition}
\newtheorem{eg}[thm]{Example}
\newtheorem{subtitle}[thm]{}
\newtheorem{ex}{Exercise}[section]
\numberwithin{equation}{section}

\def\a{\alpha}
\def\b{\beta}

\def\D{\triangle}

\def\g{\gamma}
\def\G{\Gamma}
\def\K{\nabla}
\def\l{\lambda}

\def\n{\,\vert\,}
\def\o{\theta}

\def\cb{{\mathcal{B}}}

\def\cg{{\mathcal{G}}}

\def\cl{{\mathcal{L}}}
\def\cm{{\mathcal{M}}}
\def\cn{{\mathcal{N}}}
\def\co{{\mathcal{O}}}

\def\li{\langle}
\def\ri{\rangle}
\def\n{\ \vert\ }
\def\tr{{\rm tr}}
\def\bs{\bigskip}
\def\ms{\medskip}

\def\ni{\noindent}
\def\ti{\tilde}
\def\p{\partial}

\def\Im{{\rm Im\/}}
\def\I{{\rm I\/}}

\def\diag{{\rm diag}}

\def\A{\mathbb{A}}
\def\C{\mathbb{C}}

\def\R{\mathbb{R} }

\newcommand{\beq}{\begin{equation}}
\newcommand{\eeq}{\end{equation}}
\newcommand{\beg}{\begin{eg}}
\newcommand{\eeg}{\end{eg}}
\newcommand{\bthm}{\begin{thm}}
\newcommand{\ethm}{\end{thm}}
\newcommand{\bprop}{\begin{prop}}
\newcommand{\eprop}{\end{prop}}
\newcommand{\bcor}{\begin{cor}}
\newcommand{\ecor}{\end{cor}}
\newcommand{\blem}{\begin{lem}}
\newcommand{\elem}{\end{lem}}
\newcommand{\bca}{\begin{cases}}
\newcommand{\eca}{\end{cases}}
\newcommand{\brem}{\begin{rem}}
\newcommand{\erem}{\end{rem}}
\newcommand{\bpm}{\begin{pmatrix}}
\newcommand{\epm}{\end{pmatrix}}
\newcommand{\bbm}{\begin{bmatrix}}
\newcommand{\ebm}{\end{bmatrix}}
\newcommand{\bvm}{\begin{vmatrix}}
\newcommand{\evm}{\end{vmatrix}}
\newcommand{\bdefn}{\begin{defn}}
\newcommand{\edefn}{\end{defn}}
\newcommand{\bsub}{\begin{subtitle}}
\newcommand{\esub}{\end{subtitle}}
\newcommand{\bex}{\begin{ex}}
\newcommand{\eex}{\end{ex}}
\newcommand{\ben}{\begin{enumerate}}
\newcommand{\een}{\end{enumerate}}

\def\rd{{\rm \/ d\/}}

\def\sech{{\rm sech\/}}

\def\an2{\hat A_{2n}^{(2)}}
\def\bn1{\hat B_n^{(1)}}
\def\Ker{{\rm Ker\/}}

\def\B{\mathbb{B}}
\def\bbn1{\hat \B_n^{(1)}}
\def\A{\mathbb{A}}
\def\ban2{\hat \A_{2n}^{(2)}}

\def\bh{\backslash}
\def\r0{\R^n \backslash \{0\}}

\def\bu{\bullet}
\def\n{\, | \,}
\begin{document}

\title{Darboux Transforms for the $\bn1$- hierarchy} \today

\author{Chuu-Lian Terng}
\address{Department of Mathematics\\
University of California at Irvine, Irvine, CA 92697-3875.  Email: cterng@math.uci.edu}
\author{Zhiwei Wu$^*$}\thanks{$^*$Research supported in part by NSF of China under Grant No. 11401327, and The Project Sponsored by the Scientific Research Foundation for the Returned Overseas Chinese Scholars, State Education Ministry\/}
\address{School of Mathemetics (Zhuhai)\\ Sun Yat-sen University\\ Zhuhai, Guangdong, 519082 , China. Email: wuzhiwei3@mail.sysu.edu.cn}

\maketitle

\section{Introduction}\label{iq}

Drinfeld-Sokolov in \cite{DS84} associated to each affine Kac-Moody algebra $G$ a $G$-hierarchy of soliton equations and constructed a $G$-KdV hierarchy on a cross section of certain gauge action  by pushing down the $G$-hierarchy along the gauge orbits to the cross section. 

Note that $\hat B_1^{(1)}$ is isomorphic to $\hat A_1^{(1)}$ and the  $\hat B_1^{(1)}$-KdV hierarchy is the same as the $\hat A_1^{(1)}$-KdV hierarchy with the KdV equation
\beq\label{kdv}
q_t= q_{xxx}- 3qq_x
\eeq
as the third flow. So the KdV hierarchy has two Lax representations. One is $sl(2,\R)$-valued and its natural generalization to higher rank is the $\hat A_n^{(1)}$-KdV hierarchy, i.e., the Gelfand-Dickey (GD$_n$-) hierarchy on the space of order $n$ linear differential operators on the line (cf. \cite{DS84}, \cite{TWb}) .  For example, the second $\hat A_2^{(1)}$-KdV flow is 
\beq\label{jt}
\begin{cases}
(u_1)_t = (u_1)_{xx}-\frac{2}{3}(u_2)_{xxx}+\frac{2}{3}u_2(u_2)_x,\\
(u_2)_t = -(u_2)_{xx}+2(u_1)_x.
\end{cases}
\eeq 
The other Lax representation of the KdV  is $o(2,1)$-valued, and its natural generalization to higher rank is the $\bn1$-KdV hierarchy.  

We constructed in \cite{TWd} a cross section of the gauge action for the $\bn1$ case and wrote down the $\bn1$-KdV flows as flows on this cross section. This cross section is $C^\infty(\R, V_n)$, where $V_n= \oplus_{i=1}^n \R \b_i$ and $\b_i$'s are defined by \eqref{cj}.  For example,  the third $\hat B_2^{(1)}$-KdV flow is 
\beq\label{ef}
\begin{cases}
(u_1)_t=-\frac{1}{2}(u_1)_{xxx}+\frac{3}{2}u_1(u_1)_x+3(u_2)_x,\\
(u_2)_t=(u_2)_{xxx}-\frac{3}{2}u_1(u_2)_x
\end{cases}
\eeq
for $u= u_1\b_1+ u_2\b_2$.

The $G$-KdV flows also arise naturally in geometric curve flows. For example, the flows of the central affine curvature of central affine curve flows in $\R^{n+1}\bh 0$ are the $\hat A_n^{(1)}$-KdV flows (cf. \cite{CIM13}, \cite{UP95} and \cite{TWb}), and the flows of the isotropic curvature of B-type isotropic curve flows on $\R^{n+1,n}$ (invariant under the group $O(n+1,n)$) are the $\bn1$-KdV flows (cf. \cite{TWd}). 

Next we review some basic facts about isotropic curve flows: A curve $\g$ in $\R^{n+1,n}$ is {\it isotropic\/} if $\g, \ldots, \g_x^{(n-1)}$ span a maximal isotropic subspace in $\R^{n+1,n}$ and $\g, \ldots, \g_x^{(2n)}$ are linearly independent. 
The following were proved in \cite{TWd}:
\ben
\item[(a)] There exists a parameter for an isotropic curve $\g$ in $\R^{n+1,n}$ unique up to translation such that $\li \g_x^{(n)}, \g_x^{(n)}\ri \equiv 1$ and a unique $g:\R\to O(n+1,n)$ such that $g^{-1}g_x= b+u$ for some $u\in C^\infty(\R, V_n)$ (this is the cross section of the gauge action and also is the phase space of the $\bn1$-KdV flows), where $b=\sum_{i=1}^{2n} e_{i+1,i}$ ($g$ and $u$ are called the {\it isotropic moving frame\/} and {\it isotropic curvature\/} along $\g$). 
\item[(b)]  Set 
\beq\label{jc}
\cm_{n+1,n}=\{\g:\R\to \R^{n+1,n}\,\n\, \g \, {\rm is \, isotropic,\,} \li \g_x^{(n)}, \g_x^{(n)}\ri \equiv 1\}.
\eeq
The third isotropic curve flow of B-type on $\cm_{2,1}$ is 
\beq\label{ma}
\g_t= q_x\g - q\g_x= \g_{xxx} - 3 q\g_x,
\eeq
 where $u= q(e_{12}+ e_{23})$ is the isotropic curvature. If $\g(x,t)$ is a solution of \eqref{ma}, then $q$ is a solution of the KdV \eqref{kdv}.
\item[(b)] The third isotropic curve flow of B-type on $\cm_{3,2}$ is 
\beq\label{jb}
\g_t= \g_{xxx}-u_1\g_x.
\eeq  
If $\g$ is a solution of \eqref{jb} then its isotropic curvature $u= u_1 (e_{23}+ e_{34})+ u_2(e_{14}+ e_{25})$ is a solution of the thrid $\hat B_2^{(1)}$-KdV flow \eqref{ef}. 
\een

BT for the KdV equation was constructed in \cite{WE}, and BT for the $\hat A_{n-1}^{(1)}$-KdV hierarchy was constructed in \cite{A81} and \cite{TWa}.
 In this paper, we use the loop group factorization method given in \cite{TU00} to construct Darboux transforms (DTs) and B\"acklund transformations (BTs) for the $\bn1$-KdV flows,
  then use these transforms to construct explicit soliton solutions for the $\bn1$-, $\bn1$-KdV, and isotropic curve flows of B-type. In particular, we wrote down explicit $1$-soliton solutions for the third $\hat B_2^{(1)}$-KdV flow \eqref{ef}, and for isotropic curve flows \eqref{ma} and \eqref{jb}.

 We note that the classical BT of the KdV depends on one parameter and if we apply their BT to the trivial solution $q=0$ of the KdV, we obtain 1-soliton solutions. The BTs and DTs for the KdV of this paper depend on two parameters $\a_1, \a_2$. We will see in section \ref{mu} that when we apply our DT with $\a_1>0$ and $\a_2=0$ to the trivial solution $q=0$ of the KdV we obtain $1$-soliton solutions, and we obtain $2$-soliton solutions if we apply DT to $q=0$ with $\a_1>\a_2>0$.

This paper is organized as follows: We review the construction of the $\bn1$ and $\bn1$-KdV hierarchies in section \ref{hb}. We construct DTs in section \ref{ip},  a Permutability formula in section \ref{ipa}, and scaling transforms and its relation to DTs  for the $\bn1$-flows in section \ref{mt}. In the last section, we use DTs of the $\bn1$-flows to construct explicit soliton solutions for the $\bn1$-, the $\bn1$-KdV flows, and the isotropic curve flows of B-type.

\section{The $\bn1$- and $\bn1$-KdV hierarchies}\label{hb}

Let $\cl$ be a loop algebra, $\cl_+, \cl_-$ subalgebras of $\cl$ such that $\cl= \cl_+\oplus \cl_-$ as linear subspaces, and $\{J_j\n j\geq 1\}$ a commuting linearly independent sequence in $\cl_+$ satisfying the condition that $J_j$'s are generated by $J_1$ for all $j\geq 2$. It is known that (cf. \cite{DS84}, \cite{TU00}) there is a soliton hierarchy of flows on $C^\infty(\R, Y)$, where $Y=[J_1,\cl_-]_+$ and $\xi_+$ is the projection of $\xi\in \cl$ onto $\cl_+$ along $\cl_-$.  
In this section, we review the construction of the $\bn1$- and the $\bn1$-KdV hierarchies (cf. \cite{DS84}, \cite{TWd}) using this method.

\ms
\ni{\bf The $\bn1$-hierarchy}\
 
Let  
\beq\label{am}
 \rho_n=\sum_{i=1}^{2n+1}(-1)^{n+i-1}e_{i, 2n+2-i}, 
 \eeq
and 
\beq\label{ad}
\li X, Y \ri= X^t\rho_n Y,
\eeq
the index $n$ non-degenerate bilinear form on $\C^{2n+1}$.
Note that $$\rho_n^2=\I_{2n+1}.$$ 
 Let $O_{\C}(n+1, n)$ be the group of linear isomorphisms of $\C^{2n+1}$ that preserve
$\langle \ , \ \rangle$, i.e.
\beq\label{bw}
O_{\C}(n+1, n)=\{g \in SL(2n+1, \C) \mid  g^t\rho_ng=\rho_n\}.
\eeq
Its Lie algebra is  $o_{\C}(n+1, n)=\{A \in sl(2n+1, \C) \mid  A^t\rho_n+\rho_nA=0\}$. 
Note that $A=(A_{ij})\in o_\C(n+1, n)$ if and only if 
\ben
\item[(i)]
$A_{ij}$'s are symmetric (skew-symmetric resp.) with respect to the skew diagonal line $i+j= 2n+2$ if $i+j$ is odd (even resp.),
\item[(ii)] $A_{ij}=0$ if $i+j= 2n+2$. 
\een
Let $O(n+1,n)=\{g\in O_\C(n+1, n) \n \bar g= g\}$. Then
$o(n+1, n) = \{A\in o_\C(n+1, n)\n \bar A= A\}$.
Let $\cb_n^+$ and $\cn_n^+$ denote the subalgebra of $o(n+1,n)$ of upper triangular and strictly upper triangular matrices, $B_n^+$ and  $N_n^+$ the connected subgroups of $O(n+1,n)$ corresponding to $\cb_n^+$ and $\cn_n^+$  respectively.  

Let  $\bn1$ be the Lie algebra of formal power series $\xi(\l)=\sum_{i\geq n_0} \xi_i\l^i$ with some integer $n_0$ that satisfy 
\beq\label{by}
\rho_n \xi(\l)+ \xi(\l)^t \rho_n=0, \quad \overline{\xi(\bar{\l})}=\xi(\l).
\eeq
 Let $(\bn1)_+$, $(\bn1)_-$ be the sub-algebras of $\bn1$ defined by
\begin{align*}
& (\bn 1)_+ = \{\xi(\l)=\sum_{i \geq 0} \xi_i \l^i \in \bn 1\},\\
& (\bn 1)_-=\{\xi(\l)=\sum_{i < 0}\xi_i\l^i \in \bn 1\}. 
\end{align*}
Note that 
\ben
\item[(i)] $\xi(\l)= \sum_{i\geq n_0} \xi_i\l^i \in  \bn1$ if and only if $\xi_i\in o(n+1,n)$ for all $i$, 
\item[(ii)] $\bn 1=(\bn 1)_+\oplus (\bn 1)_-$ is a direct sum of linear subspaces.
\een
Henceforth, for $\xi(\l)=\sum_i \xi_i\l^i$, we use the following notations:
\beq\label{ax}
\xi_+(\l)= \sum_{i\geq 0} \xi_i \l^i, \quad \xi_-(\l) = \sum_{i<0} \xi_i \l^i.
\eeq

Let 
\beq\label{dc}
J_B(\l) =\b \l + b,
\eeq
where
\beq\label{db}
\b=\frac{1}{2}(e_{1, 2n}+e_{2, 2n+1}), \quad b= \sum_{i=1}^{2n}e_{i+1, i}.
\eeq
Note that $J_B^{2j}\not\in \bn1$, $J_{B}^{2j-1} (j \geq 1) \in (\bn 1)_+$, and
\beq\label{dh}
 J_B^{2n+1}(\l)=\l J_B(\l),
\eeq

\bthm\label{zc}(\cite{DS84}, \cite{TWd}) Let $q\in C^\infty(\R, \cb_n^+)$, and $J_B(\l)= b+ \b\l$ as in \eqref{dc}. Then
\ben
\item there exists $M(q,\l)(x)=\I_{2n+1} +\sum_{i<0} M_i(x)\l^i$ satisfying 
\beq\label{bu}
f(\l)^t\rho_nf(\l)=\rho_n, \quad \overline{f(\bar \l)}=f(\l),
\eeq 
 and  
\beq\label{bb}
M^{-1}(q,\l)(\p_x+J_B)M(q,\l)= \p_x+J_B + q,
\eeq
\item $T(q,\l):= M^{-1}(a,\l)J_B(\l) M(q,\l)$ is in $\bn1$ and satisfies 
\beq\label{ch}
\bca
[\p_x+\b\l + b+q, T(q, \l)]=0,\\
T^{2n+1}(q,\l)= \l ( \b\l+b),
\eca
\eeq
\item \eqref{ch} has a unique solution of the form 
$$T(q,\l)= \b\l + \sum_{i\leq 0} T_{1,i}(q) \l^i,$$
where $T_{1, i}(q)$ is a differential polynomial in $q$ for $i\leq 0$.
\een
\ethm

Expand $T^{2j-1}(q,\l)$ as a power series in $\l$, 
\beq\label{dg}
T^{2j-1}(q, \l)=\sum_{i \leq [\frac{2j-1}{2n+1}]+1}T_{2j-1, i}(q)\l^i.
\eeq

Note that if $A, B$ lie in an associative algebra satifying $[A,B]=0$, then $[A, B^i]=0$ for all $i\geq 1$. So
the first equation of \eqref{ch} implies that 
$$[\p_x+J_B(\l) + q, T^{2j-1}(q,\l)]=0.$$
Compare coefficient of $\l^i$ of the above equation to get
\beq\label{bai}
[\p_x+b+q, T_{2j-1, i}(q)]= [T_{2j-1, i-1}(q), \b].
\eeq
In particular for $i=0$, we obtain
\beq\label{ba}
[\p_x+b+q, T_{2j-1,0}(q)]= [T_{2j-1, -1}(q), \b].
\eeq
So 
\beq\label{ah}
q_t= [\p_x+b+q, T_{2j-1,0}(q)]
\eeq is a flow on $C^\infty(\R, Y_n)$, where 
$$Y_n= [J_B, (\bn1)_-]_+= [\b, o(n+1,n)].$$ 
Since $Y_n\subset \cb_n^+$, the flow \eqref{ah} can be viewed as a flow on $C^\infty(\R, \cb_n^+)$. We call the flow \eqref{ah} the {\it $(2j-1)$-th flow of the $\bn1$-hierarchy\/} on $C^{\infty}(\R, \cb_n^+)$. 
Note that  if $q=(q_{ij})$ is a solution of \eqref{ah}, then $q_{ij}(x,t)= q_{ij}(x,0)$ if $i<2n$ and $j>2$. 

\ms
\ni {\bf The $\bn1$-KdV hierarchy}\
 
 \ms

The group $C^\infty(\R, N_n^+)$ acts on $C^\infty(\R, \cb_n^+)$ by gauge transformations,
$$\D(\p_x+b+q)\D^{-1}= \p_x+ b+ \D\ast q,$$
 where  $q\in C^\infty(\R, \cb_n^+)$, $\D\in C^\infty(\R, N_n^+)$, and
\beq\label{dr}
 \D\ast q= \D(b+q)\D^{-1}-\D_x\D^{-1}.\eeq 
 Note that  
 \beq\label{dq}
\D(\p_x+J_B(\l)+ q)\D^{-1}=\p_x+J_B(\l)+ \D \ast q,
\eeq

The following Theorem gives a cross section of this action:
 
\bthm\label{mj} (\cite{TWd})  Let $V_n= \oplus_{i=1}^n \R \b_i$, where
\beq\label{cj}
\b_i= e_{n+1-i, n+i} + e_{n+2-i, n+1+i}, \quad 1\leq i\leq n.
\eeq
Given $q \in C^{\infty}(\R, \cb_n^+)$, then there
exist a unique $\D \in C^{\infty}(\R, N_n^+)$ and $u\in C^\infty(\R, V_n)$ such that
\beq\label{aa}
\D(\p_x+b+q)\D^{-1}=\p_x+b+u.
\eeq
 Moreover, entries of $\D$ and $u$ can be computed from \eqref{aa} and are differential polynomials of $u$ and $q$.
\ethm

\bthm\label{mja} (\cite{TWd}) If $u\in C^\infty(\R, V_n)$, then there exists a unique linear differential operator $P_u:C^\infty(\R, V_n^t)\to C^\infty(\R, o(n+1,n))$ satisfying
\beq\label{kd}
\bca [\p_x+b+u, P_u(v)]\in C^\infty(\R, V_n),\\ \pi_0(P_u(v))=v,\eca
\eeq
where $\pi_0:o(n+1,n)\to V_n^t$ is the projection defined by $$\pi_0(y)=\sum_{i=1}^{n}y_{n+i, n+1-i}\b_i^t, \quad {\rm where\,\,} y=(y_{ij}).$$ Moreover, coefficients of $P_u(v)$ are polynomial differential of $u$. 
\ethm

\bthm\label{mk} (\cite{TWd}) If $u\in C^\infty(\R, V_n)$, then
\beq\label{hz}
\eta_j(u):= T_{2j-1,0}(u)- P_u(\pi_0(T_{2j-1,0}(u)))
\eeq
is a $\cn_n^+$-valued differential polynomial of $u$. 
\ethm

\bdefn The $(2j-1)$-th $\bn1$-KdV flow is 
\beq\label{ai}
u_t= [\p_x+ b+u, P_u(\pi_0(T_{2j-1,0}(u)))]
\eeq
 for $u:\R^2\to V_n$. 
\edefn

The next theorem states that the flow obtained by pushing down the $\bn1$-hierarchy along the gauge orbit of $C^\infty(S^1, N_n^+)$ is the $\bn1$-KdV hierarchy. 

\bthm\label{mb} (\cite{TWd})
Let $q:\R^2\to \cb_n^+$ be a smooth solution of the $(2j-1)$-th $\bn1$-flow \eqref{ah}, and $\D:\R^2\to N_n^+$ such that $u(\cdot, t):= \D(\cdot, t)\ast q(\cdot, t)$ lies in $V_n$ for all $t$. Then 
\ben
\item $u$ is a solution of \eqref{ai},
\item $\D_t\D^{-1}= \eta_j(u)$, where $\eta_j(u)$ is defined by \eqref{hz}.
\een
\ethm

\ms \ni{\bf Lax pairs and Frames}

The following is a well-known and elementary fact. 

\blem\label{mc} Let $\cg$ be the Lie algebra of $G$, and $A, B\in C^\infty(\R^2, \cg)$. Then the following linear system 
$$\bca \phi_x= \phi A,\\ \phi_t= \phi B,\eca$$
is solvable for $\phi:\R^2\to G$ if and only if 
$$A_t= [\p_x+A, B].$$
\elem

The recursive formula \eqref{bai} and Lemma \ref{mc} give the following Theorem.

\bthm\label{ce} The following statements are equivalent for smooth $q:\R^2 \to \cb_n^+$:
  \ben
  \item $q$ is a solution of the {\it $(2j-1)$-th $\bn1$-flow\/} \eqref{ah},
  \item $q_t= [\p_x+ J_B(\l)+ q, (T^{2j-1}(q,\l))_+]$,
  \item
  \beq\label{ee}
  [\p_x+ J_B(\l)+ q, \p_t +(T^{2j-1}(q,\l))_+]=0,
  \eeq
    \item
the following linear system is solvable for $F(x,t, \l)\in O_\C(n+1,n)$,
\beq\label{gsa}
\bca F^{-1}F_x= J_B(\l)+q, \\ F^{-1}F_t= (T^{2j-1}(q,\l))_+,\\ \overline{F(x,t,\bar\l)}= F(x,t,\l),\eca
\eeq
where $\xi_+$ is defined by \eqref{ax}.
\een
\ethm

It follows from \eqref{bai}, Lemma \ref{mc} and Theorem \ref{mb} that we have the following. 

\bthm\label{cek} The following statements are equivalent for smooth $u:\R^2 \to V_n$:
\ben
\item $u$ is a solution of the $(2j-1)$-th $\bn1$-KdV flow \eqref{ai},
\item $u_t=[\p_x+J_B(\l)+u, (T^{2j-1}(u,\l))_+-\eta_j(u)]$,
\item 
\beq\label{eek}
[\p_x+J_B(\l)+ u, \p_t+ (T^{2j-1}(u,\l))_+-\eta_j(u)]=0,
\eeq
\item the following linear system 
 \beq\label{en}
 \bca E^{-1}E_x= J_B(\l)+ u, \\ E^{-1}E_t= (T^{2j-1}(u,\l))_+-\eta_j(u),\\
 \overline{E(x,t,\bar\l)}=E(x,t,\l), \eca
 \eeq is solvable for $E(x,t,\l) \in O_\C(n+1,n) $,  
 \een 
 where $\eta_j(u)$ is defined by \eqref{hz}. 
 \ethm

\bdefn  \ \ben
\item We call \eqref{ee} (\eqref{eek} resp.) the {\it Lax pair\/} of solution $q$ of the $(2j-1)$-th $\bn1$- flow ($u$ of the $(2j-1)$-th $\bn1$-KdV flow resp.). 
\item We call a solution $F(x, t, \l)$ of \eqref{gsa} ($E(x,t,\l)$ of \eqref{en} resp.) a \emph{frame} of the solution $q$ of the $(2j-1)$-th $\bn1$-flow \eqref{ah} ($u$ of the $(2j-1)$-th $\bn1$-KdV flow \eqref{ai}) resp.)  if $F$ ($E$ resp.) is holomorphic for $\l\in \C$.
\een
\edefn

\beg\label{ia} (\cite{TWd})   For the case $n=1$, we have $q=q_1(e_{11}- e_{33}) + q_2 (e_{12}+ e_{23})$, $\D$ and $u$  given in Theorem \ref{mj} are
\begin{align*}
& \D=\bpm 1 & -q_1 & \frac{1}{2}q_1^2 \\ 0 & 1 & -q_1 \\ 0 & 0 & 1 \epm, \\
& u=q(e_{12}+ e_{21})= (q_2+\frac{1}{2}q_1^2+(q_1)_x) (e_{12}+ e_{23}), 
\end{align*}
 the $P_u(v)$ given in Theorem \ref{mja} is
\beq\label{dka}
P_u(v) =\bpm -(v_1)_x & -(v_1)_{xx}+q v_1 & 0  \\
v_1 & 0 & -(v_1)_{xx}+qv_1 \\ 0 & v_1 & (v_1)_x \epm,
\eeq
 the third $\hat B_1^{(1)}$-KdV is the KdV \eqref{kdv}, and the Lax pair is 
 \beq\label{bs}
 \left[\p_x+ J_B+u, \p_t+ J_B^3+\left(\begin{array}{ccc}q_x&\frac{q}{2}\l+q_{xx}-q^2&0\\-q&0&\frac{q}{2}\l+q_{xx}-q^2\\0&-q&-q_x\end{array}\right)\right]=0.
 \eeq
\eeg

As a consequence of Theorems  \ref{mb}, \ref{ce}, and \ref{cek}, we obtain the following two Propositions:

\bprop\label{cga} Let $F(x,t,\l)$ be a frame of a solution  $q$ of the $(2j-1)$-th $\bn1$-flow \eqref{ah}, and $\D:\R^2\to N_n^+$ such that $u:=\D\ast q$ lies in $V_n$, where the gauge action $\ast$ is defined by \eqref{dr}.   Then $u$ is a solution of the $(2j-1)$-th $\bn 1$-KdV flow \eqref{ai} and $E(x,t,\l)=F(x,t,\l) \D^{-1}(x,t)$ is a frame of the solution $u$ of the $\bn1$-KdV flow \eqref{ai}.
\eprop

\bprop\label{do} Let $E(x,t,\l)$ be a frame of the solution $u$ of the $(2j-1)$-th $\bn1$-KdV flow \eqref{ai}. Suppose $\D:\R^2\to N_n^+$ satisfying $\D_t\D^{-1}=\eta_j(u)$, where $\eta_j(u)$ is defined by \eqref{hz}.  Then $q(\cdot, t)= \D(\cdot, t)^{-1}\ast u(\cdot, t)$ is a solution of \eqref{ah} and $F(x,t,\l)= E(x,t,\l)\D(x,t)$ is its frame, where the action $\ast$ is defined by \eqref{dr}.  
\eprop

It follows from Example \ref{ia} and Proposition \ref{cga} that we have

\bcor \label{at} (\cite{TWd}) If $q= q_1(e_{11}-e_{33})+ q_2(e_{12}+e_{23})$ is a solution of the the $(2j-1)$-th $\hat B_1^{(1)}$-flow, then $u= (q_2+\frac{1}{2} q_1^2+ (q_1)_x)(e_{12}+e_{23})$ is a solution of the $(2j-1)$-th $\hat B_1^{(1)}$-KdV flow.
\ecor

\bs

\section{DT for the $\bn1$-flows}\label{ip}

In this section, we use the loop group factorization method given in \cite{TU00} to construct Darboux Transforms for the $\bn1$- and the $\bn1$-KdV flows. 

Let $(\hat\B_n^{(1)})_+$ denote the group of holomorphic maps $f:\C\to GL(2n+1, \C)$ satisfying  \eqref{bu}, 
 and $\R \B_n^{(1)}$ the group of rational maps $f:\C\cup \{\infty\}\to GL(n,\C)$ satisfying \eqref{bu} and $f(\infty)=\I$. Then the Lie algebras of $(\hat \B_n^{(1)})_+$ and $\R\bbn1$ are subalgebras of $(\bn1)_+$ and $(\bn1)_-$ respectively.   
 
 First we review the general method of using the loop group factorization to construct DTs given in \cite{TU00}:
 
\ben
\item[(i)] Find simple elements, i.e., elements  in $\R\bbn1$ with minimal number of poles.
\item[(ii)] Given $f\in (\bn1)_+$ and a simple element $g\in \R\bbn1$, construct the factorization $g f= \ti f \ti g$ explicitly with $\ti f\in (\bn1)_+$ and $\ti g$ a simple element in $\R\bbn1$. 
\item[(iii)] Solve the linear system given by the Lax pair of a given solution of the soliton hierarchy to get $F(x,t,\cdot)\in (\bn1)_+$, i.e., construct a {\it frame\/} of the given solution. 
\item[(iv)] Given a simple element $g$ and a frame $F(x,t,\l)$ of a solution $q$, we use (ii) to factor 
$$g(\l)F(x,t,\l)= \ti F(x,t,\l) \ti g(x,t,\l)$$ with $g(x,t, \cdot)$ a simple element and $\ti F(x,t,\cdot)$ in the positive loop group for each $(x,t)$. It was proved in \cite{TU00} that $\ti F$ is a frame of a new solution of the soliton hierarchy.
\item[(v)] The rational loop group acts on the space of solutions, DTs are given by rational loops with minimal number of poles. 
\item[(vi)] Permutability formulas arise from (v) and relations among simple elements.
\een 

The next Proposition proves the uniqueness of factorization.

\bprop\label{ho} If $f_1 f_2= g_1 g_2$ with $f_1, g_1\in (\bbn1)_+$ and $f_2, g_2\in \R\bbn1$, then $f_1=g_1$ and $f_2= g_2$. 
\eprop

\begin{proof}
Note that $h:=g_1^{-1}f_1= g_2 f_2^{-1}$. The left hand side is holomorphic for $\l\in \C$ and the right hand side is holomorphic at $\l=\infty$. So it is holomorphic in $\C\cup \{\infty\}$. Therefore $h$ is a constant. But the right hand side at $\l=\infty$ is equal to $\I$. So $h=\I$. 
\end{proof}

Note that if $F(x,t,\l)$ is a frame of a solution $q$ of the $(2j-1)$-th $\bn1$-flow \eqref{ah}, then $F(x,t,\cdot)$ is in $(\bbn1)_+$ for all $(x,t)\in \R^2$.  

\bthm\label{gx}(\cite{TU00}) Let $F(x,t,\l)$ be a frame of a solution $q$ of the $(2j-1)$-th  $\bn1$-flow \eqref{ah}, and $f\in \R\bbn1$. Assume that we can factor 
$$f(\l) F(x,t,\l) = \ti F(x,t,\l) \ti f(x,t,\l)$$
such that $\ti F(x,t,\cdot)\in (\bbn1)_+$ and $\ti f(x,t,\cdot)\in \R\bbn1$. Then 
\beq\label{bz}
\ti q= q+[\ti f_{-1}, \b],
\eeq 
is a new solution of \eqref{ah}, and $\ti F$ is a frame of $\ti q$, where $\ti f_{-1}$ is the coefficient of $\l^{-1}$ of $\ti f$ as a power series of $\l$.    
 \ethm

\bdefn  Let $F(x,t,\l)$ be the frame of a solution $q$ of the $(2j-1)$-th $\bn1$-flow satisfying $F(0,0,\l)=\I_{2n+1}$, and $f\in \R\bbn1$.  We use $f\bu q$ to denote the solution $\ti q$ defined by \eqref{bz} from $q$, $F$ and $f$. 
\edefn

\bcor\label{gxa} (\cite{TU00})
Let $f_1, f_2\in \R\bbn1$, and  $q$ a solution of the $(2j-1)$-th $\bn1$-flow \eqref{ah}. If $f_2\bu q$ and $f_1\bu(f_2\bu q)$ exist, then $(f_1f_2)\bu q= f_1\bu (f_2\bu q)$.  
\ecor

Next we construct simple elements in $\R\bn1$. 
We say $\pi$ is {\it a projection of $\R^{n+1,n}$ onto $V_1$ along $V_2$\/} if  $\pi^2=\pi$, $\Im\pi= V_1$, and $\Ker\pi= V_2$.
 Note that if $\pi$ is the projection onto $V_1$ along $V_2$, then 
  $$\pi^\perp:= \I_{2n+1}-\pi$$
 is the projection onto $V_2$ along $V_1$. 

\bprop\label{cn} (\cite{TU00})   Let $\a_1 \neq \a_2 \in \R$, $\pi$ a projection of $\R^{2n+1}$, and 
\beq\label{ct}
h_{\a_1, \a_2, \pi}=\I_{2n+1}+\frac{\a_1-\a_2}{\l-\a_1}\pi^\perp = \frac{\l-\a_2}{\l-\a_1}\I_{2n+1} -\frac{\a_1-\a_2}{\l-\a_1} \pi.
\eeq 
Then we have the following:
\ben
\item $h_{\a_1,\a_2,\pi}^{-1}= h_{\a_2,\a_1, \pi}$.
\item Let $f:\C\to SL(2n+1,\C)$ be a  holomorphic map satisfying $\overline{f(\bar\l)}= f(\l)$, and 
$$\ti V_1= f^{-1}(\a_1)(\Im \pi), \quad \ti V_2= f^{-1}(\a_2)(\Ker \pi).$$  If $\ti V_1\cap \ti V_2=\{0\}$, then 
\ben
\item $\R^{n+1,n}= \ti V_1\oplus \ti V_2$,
\item let $\ti\pi$ be the projection of $\R^{n+1,n}$ onto $\ti V_1$ along $\ti V_2$, then $$\ti f= h_{\a_1,\a_2,\pi} f h_{\a_1,\a_2,\ti\pi}^{-1}$$ is homomorphic for $\l\in \C$ and $\overline{\ti f(\bar\l)}=\ti f(\l)$.
\een
\een
\eprop

\bcor\label{ijb} Let $\a_1,\a_2,\pi, f, \ti f$ be as in Proposition \ref{cn}. Then
\begin{align}
\ti f^{-1}(\a_1)&= f^{-1}(\a_1)\pi + \ti\pi^\perp f^{-1}(\a_1)\pi^\perp +(\a_1-\a_2)\ti\pi^\perp(f^{-1})_\l(\a_1) \pi.\label{co1}\\
\ti f^{-1}(\a_2)&= \ti \pi f^{-1}(\a_2)+ \ti \pi^\perp f^{-1}(\a_2)\pi^\perp -(\a_1-\a_2)\ti \pi (f^{-1})_\l(\a_2) \pi^\perp. \label{co}
\end{align}
\ecor

\begin{proof} Use the Taylor series expansion of $f^{-1}(\l)$ and $(\l-\a_2)^{-1}$ at $\l=\a_1$ to compute $\ti f^{-1}(\a_1)$. Similar computation gives the formula for $\ti f^{-1}(\a_2)$.
\end{proof}

The following two Lemmas can be proved using elementary linear algebra.

\blem\label{cs}\

\ben 
\item Let $A\in gl(2n+1,\C)$, and $A^\sharp$ the adjoint of $A$ with respect to the bilinear form $\li\, , \ri$ defined by \eqref{ad}, i.e.,  $\li AX, Y\ri = \li X, A^\sharp Y\ri$ for all $X, Y\in \C^{2n+1}$. Then
\beq\label{ir}
A^\sharp= \rho_n A^t \rho_n.
\eeq 
\item Let $\R^{n+1, n}= V_1\oplus V_2$, and $V_i^\perp=\{v\in \R^{n+1, n}\n  \li v, V_i\ri =0\}$, $i=1, 2$.   If $\pi$ is the projection of $\R^{n+1, n}$ onto $V_1$ along $V_2$, then its adjoint $\pi^\sharp$ is the projection of $\R^{n+1, n}$ onto $V_2^{\perp}$ along $V_1^\perp$. 
 \een
\elem

\blem \label{bc}  Let $\pi$ be the projection onto $V_1$ along $V_2$, and $\pi^\sharp=\rho_n\pi^t \rho_n$ the adjoint of $\pi$ be as in Lemma \ref{cs}.  Then 
\beq\label{cua}
\pi \pi^{\sharp}=\pi^{\sharp}\pi={\bf 0}
\eeq
 if and only if
\beq\label{cu}
V_1 \subset V_1^{\perp}, \quad V_2^{\perp} \subset V_2.
\eeq
\elem

Use the above two lemmas and a direct computation to get the following.

\bprop\label{bca}
Let $\a_1\not=\a_2\in \R$, and $\pi$ a projection of $\R^{n+1,n}$ satisfying \eqref{cua}. Then
\beq\label{bd}
g_{\a_1, \a_2, \pi}=h_{\a_2, \a_1, \pi^{\sharp}}h_{\a_1, \a_2, \pi}=\I_{2n+1}+\frac{\a_2-\a_1}{\l -\a_2}\pi+\frac{\a_1-\a_2}{\l-\a_1}\pi^{\sharp}
\eeq
 is in $\R\bbn1$ and 
\beq\label{bda}
g_{\a_1, \a_2, \pi}^{-1}= g_{\a_1, \a_2, \pi^\sharp}.
\eeq
\eprop

The next Proposition gives a necessary and sufficient condition on projections $\pi$ and $\o$ so that
$h_{\a_2,\a_1, \o} h_{\a_1, \a_2,\pi}$ is in $\R \bbn1$.

\bprop\label{dd} Given $\a_1\not=\a_2\in \R$ and projections $\o, \pi$.$h_{\a_2,\a_1, \o} h_{\a_1, \a_2,\pi}$ satisfies \eqref{bu} if and only if $\o$ and $\pi$ satisfy
\beq\label{cz}
\bca
\o\pi^\perp = (\o^\sharp)^\perp \pi^\sharp,\\ 
\o^\perp \pi= \o^\sharp (\pi^\sharp)^\perp
\eca
\eeq
\eprop

Next we construct the factorization of $g_{\a_1,\a_2,\pi} f$ for $f\in (\bbn1)_+$. First we need a Lemma. 
 
\blem \label{gb}  Assume that $f:\C\to GL(2n+1, \C)$ is meromorphic and satisfying $\overline {f(\bar \l)}=f(\l)$. If $f=f_1f_2$ such that $f_1$ is holomorphic on $\C$ and $f_2$ is rational with $f_2(\infty)=\I_{2n+1}$.  If $\overline{f_i(\bar \l)}=f_{i}(\l)$ for $i=1, 2$, and $f$ satisfies \eqref{bu}. Then $f_1 \in (\bbn1)_+$ and $f_2 \in \R\bbn1$.
\elem

\begin{proof} Let $g_i(\l)=\rho_n(f_i(\l)^t)^{-1}\rho_n^{-1}$ for $i=1, 2$. Then $\overline{g_i(\bar \l)}=g_i(\l)$. Moreover, $g_1(\l)$ is holomorphic in $\C$ and $g_2(\l)$ rational with $g_2(\infty)=\rho_n(f_2(\infty)^t)^{-1}\rho_n^{-1}=\I_{2n+1}$. Note that 
$$g_1g_2=\rho_n(f_1^t)^{-1}(f_2^t)^{-1}=\rho_n((f_1f_2)^t)^{-1}\rho_n=f_1f_2.$$
Then by Proposition \ref{ho}, $g_1=f_1$ and $g_2=f_2$. 
\end{proof}

Lemma \ref{bc} and Proposition \ref{dd} imply that if $\pi$ satisfies \eqref{cua}, then $h_{\a_2,\a_1, \pi^\sharp}h_{\a_2,\a_2, \pi}$ is in $\R \bbn1$. Let $f\in (\bbn1)_+$.  By Proposition \ref{cn}, we have
$$h_{\a_2,\a_1,\pi^\sharp} h_{\a_1,\a_2, \pi} f= h_{\a_2,\a_1, \pi^\sharp} \hat f h_{\a_1, \a_2, \ti\pi},$$
where $\ti \pi$ is the projection onto $f^{-1}(\a_1)(\Im\pi)$ along $f^{-1}(\a_2)(\Ker\pi)$.  Apply Proposition \ref{cn} again to get 
 $$h_{\a_2,\a_1, \pi^\sharp} \hat f= \ti f h_{\a_2,\a_1, \o},$$
 where $\o$ is the projection onto $W_2=\hat f^{-1}(\a_2)(\Ker\pi^\perp)$ along $$W_1=\hat f^{-1}(\Im\pi^\perp).$$ Thus we have 
 $$h_{\a_2,\a_2,\pi^\sharp} h_{\a_1,\a_2,\pi}f = \ti f h_{\a_2,\a_1,\o} h_{\a_1,\a_2,\ti\pi},$$
 where $\ti f$ is in $(\bbn1)_+$. 
Note that we need to use Corollary \ref{ijb} to compute $W_1$ and $W_2$. It is not clear that $\o= \ti\pi^\sharp$, i.e., $W_i=\ti V_i^\perp$. We are able to prove this is true when the rank of $\pi$ is one (the proof is long).

\bthm\label{ij} Let $\a_1\not=\a_2$ be real constants, and $\pi$ a rank one projection of $\R^{n+1,n}$ onto $V_1$ along $V_2$ satisfying \eqref{cua}. Let $f$ be a meromorphic map that is holomorphic at $\l=\a_1, \a_2$ and satisfies \eqref{bu}. Set  
$$\ti V_i = f(\a_i)^{-1}(V_i), \quad i=1,2.$$ If  $\ti V_1\cap \ti V_2=\{0\}$, then we have the following:
\ben 
\item  $\R^{n+1,n}= \ti V_1\oplus \ti V_2$, and the projection  $\ti\pi$ onto $\ti V_1$ along $\ti V_2$ satisfies \eqref{cua}.
\item $\ti f:= g_{\a_1,\a_2, \pi} f g_{\a_1,\a_2,\ti\pi}^{-1}$ is holomorphic at $\l=\a_1, \a_2$, where $g_{\a_1, \a_2,\pi}$ is defined by \eqref{bd}.
\een 
\ethm

\begin{proof} 
 Since $f$ satisfies \eqref{bu} and $\a_i\in \R$, $f(\a_i)\in O(n+1, n)$. Hence $f(\a_i)^{-1}(V_i^\perp)= \ti V_i^\perp$ and $\ti V_1, \ti V_2$ satisfy \eqref{cu}. This proves (1).

By Proposition \ref{cn},  $k:= h_{\a_1,\a_2,\pi} f h_{\a_1,\a_2, \ti\pi}^{-1}$ is holomorphic at $\l=\a_1, \a_2$. Let  
\beq\label{mv}
W_i= k^{-1}(\a_i)(V_i^\perp), \quad i=1,2.
\eeq
Use Proposition \ref{cn} again to see that 
\beq\label{da}
\ti f:= h_{\a_2,\a_1,\pi^\sharp} k h_{\a_2, \a_1, \o}^{-1}
\eeq is holomorphic at $\l=\a_1, \a_2$, where $\o$ is the projection onto $W_2$ along $W_1$,  So we have 
\begin{align*}
g_{\a_1, \a_2, \pi}f &= h_{\a_2,\a_1, \pi^\sharp} h_{\a_1,\a_2, \pi} f, \quad {\rm by\, (2),} \\
&= h_{\a_2,\a_1, \pi^\sharp} k h_{\a_1, \a_2,\ti\pi}, \quad {\rm by \, \eqref{da},} \\
&= \ti f h_{\a_2, \a_1, \o} h_{\a_1,\a_2, \ti\pi},
\end{align*}
 $\ti f$ is holomorphic at $\l=\a_1, \a_2$. and $\overline{\ti f(\bar \l)}=\ti f(\l)$. By Lemma \ref{gb}, $h_{\a_2, \a_1, \theta}h_{\a_1, \a_2, \ti\pi}$ satisfies \eqref{bu}. 
 
  Note that $\ti\pi f^{-1}(\a_2)(V_2^\perp)= \ti\pi \ti V_2^\perp$. By (1), $\ti V_2^\perp\subset \ti V_2$. So $\ti\pi f^{-1}(\a_2)(V_2^\perp)=0$. It follows from Corollary \ref{ijb} that the $W_1, W_2$ defined by \eqref{mv} are given by
  \begin{align*}
& W_1=(f^{-1}(\a_1)\pi+\ti \pi^{\perp}f^{-1}(\a_1)\pi^{\perp}+(\a_1-\a_2)\ti \pi^{\perp}(f^{-1})_\l(\a_1)\pi) V_1^{\perp}, \\
& W_2=(\ti \pi^{\perp}f^{-1}(\a_2)\pi^{\perp}-(\a_1-\a_2)\ti \pi (f^{-1})_\l(\a_2)\pi^{\perp})V_2^{\perp},
\end{align*}

 Since $h_{\a_2, \a_1, \theta}h_{\a_1, \a_2, \ti\pi}$ satisfies \eqref{bu},   Proposition \ref{dd} implies that   
\beq\label{de}
\o \ti\pi^\perp = (\o^\sharp)^\perp \ti\pi^\sharp, \quad \o^\perp \ti\pi= \o^\sharp (\ti\pi^\sharp)^\perp.
\eeq
We claim that  $\theta=\ti \pi^\sharp$, i.e., $W_i= \ti V_i^\perp$ for $i=1,2$. 

First we prove $W_2=\ti V_2^\perp$. It follows from Corollary \ref{ijb} that we have 
\beq\label{di}
k^{-1}(\a_2)= \ti\pi f^{-1}(\a_2) +\ti\pi^\perp f^{-1}(\a_2) \pi^\perp -(\a_1-\a_2)\ti\pi (f^{-1})_\l(\a_2) \pi^\perp
\eeq
Since $V_2^\perp\subset V_2$, $\pi$ is identity on $V_2^\perp$. So we have 
\ben
\item[(i)]
$\ti\pi f^{-1}(\a_2)V_2^\perp=0$, 
\item[(ii)] 
$W_2= k^{-1}(\a_2)(V_2^\perp)\subset \ti V_2^\perp + \ti V_1$.
\een
Moreover, 
 if $w_2\in W_2$ is not zero, then
 there exist $x\in \ti V_2^\perp$ and $\xi\in\ti V_1$ such that $w_2= x+\xi$ and $x\not=0$.  

We use
\eqref{de} and
\begin{align*}
& \Im\ti\pi= \ti V_1,  \quad \Im\ti\pi^\perp= \ti V_2 \supset \ti V_2^\perp, \\
& \Im\ti\pi^\sharp= \ti V_2^\perp, \quad \Im(\ti\pi^\sharp)^\perp= \ti V_1^\perp \supset \ti V_1,
\end{align*}
to see that 
\beq\label{dea}
\theta(x)=(\theta^{\sharp})^{\perp}(x), \quad \theta^{\perp}(\xi)=\theta^{\sharp}(\xi).
\eeq
It follows from \eqref{dea} that we have 
\begin{align*}
x+\xi=\theta(x+\xi)=\theta(x)+\theta(\xi)=x-\theta^{\sharp}(x)+\xi-\theta^{\sharp}(\xi).
\end{align*}
Hence $w_2=x+\xi \in \Ker(\theta^{\sharp})=W_2^{\perp}$. This proves that $W_2 \subset W_2^{\perp}$.

It follows from $\ti V_1 \subset \ti V_1^{\perp}$, $\ti V_2^{\perp} \subset \ti V_2$ and $W_2\subset W_2^\perp$ that we have $\li x, x\ri=\li \xi,\xi\ri=\li x+\xi, x+\xi\ri=0$. So
\beq\label{kf1}
\li x, \xi \ri=0.
\eeq 
If $\xi\not=0$, then since $\dim\ti V_1 =1$, we have $\ti V_1= \R \xi$. By \eqref{kf1}, $x \in \ti V_1^{\perp}$. But by assumption, $x\in \ti V_2^\perp$, so $x=0$, a contradiction.  This proves that $\xi=0$ and $w_2=x\in \ti V_2^\perp$. Hence $W_2\subset \ti V_2^\perp$. But these two subspaces have the same dimension. So $W_2= \ti V_2^\perp$. 

To prove $W_1= \ti V_1^\perp$, it is suffices to prove $W_1^\perp= \ti V_1$. We first claim that $W_1^\perp\subset W_1$.  
Given $y \in \R^{n+1, n}$, write $y=y_1+y_2$, with $y_1\in \ti V_1$ and $y_2 \in \ti V_2$. Write $y_2=y_{21}+y_{22}$ with $y_{21} \in \ti V_1^{\perp}$ and $y_{22} \in \ti V_2^{\perp}$. Since $\ti V_2^{\perp} \subset \ti V_2$, $y_{21} \in \ti V_2$, hence $y_{21} \in \ti V_1^{\perp} \cap \ti V_2$.
We have proved $y_{21}$ and $y_{22}$ are in $\ti V_2= W_2^\perp$. So 
$$\o^\sharp(y_{21})=\o^\sharp (y_{22})=0.$$
Since $y_1\in \ti V_1\subset \ti V_1^\perp$, $(\ti\pi^\sharp)^\perp(y_1)=y_1$. 
Compute directly to get
\begin{align*}
&\o\o^\sharp(y)= \o\o^\sharp(y_1+ y_{21}+ y_{22}) = \o\o^\sharp(y_1)\\
&= \o\o^\sharp(\ti\pi^\sharp)^\perp (y_1), \quad {\rm by\,\,  \eqref{de},}\\
& = \o \o^\perp\ti\pi(y_1)= \o\o^\perp(y_1).
\end{align*}
Hence we obtain 
$$\o(\o^\sharp(y)- \o^\perp(y_1))=0,$$
which implies that $\o^\sharp (y)-\o^\perp (y_1)\in \Ker\o= W_1$.  But $\o^\perp(y_1)\in W_1$. So $\o^\sharp(y)\in W_1$, which proves the claim. 

Next we prove $W_1^\perp=\ti V_1$. Given $0\not=w_1\in W_1^\perp$,  we apply \eqref{de} to $w_1 \in W_1^{\perp}$ and get
\beq\label{kf2}
\ti \pi(w_1)=\ti \pi \theta^{\perp}(w_1)=(\ti \pi^{\sharp})^{\perp}\theta^{\sharp}(w_1)=(\ti \pi^{\sharp})^{\perp}(w_1).
\eeq
Write $w_1= z_1+z_2= \eta_1+\eta_2$ with $z_i\in \ti V_i$ and $\eta_i\in \ti V_i^\perp$. By \eqref{kf2}, we have $z_1= \eta_1$. So  $z_2=\eta_2$. This shows that $z_1\in \ti V_1\cap \ti V_1^\perp =\ti V_1$ and $z_2\in \ti V_2^\perp$. Since $\ti V_1\subset \ti V_1^\perp$, $\ti V_2^\perp \subset \ti V_2$, and $W_1^\perp\subset W_1$, we have 
$\li z_1, z_1\ri= \li z_2, z_2\ri = \li w_1, w_1\ri=0$. Hence 
\beq\label{kf3}
\li z_1, z_2\ri=0.
\eeq
 If $z_2\not=0$, then since $\dim\ti V_2^\perp= 1$, $\ti V_2^\perp=\R z_2$. By \eqref{kf3}, $z_1$ is in  $(\ti V_2^\perp)^\perp= \ti V_2$. So $z_1\in \ti V_1\cap \ti V_2=0$. This implies that $w_1= z_2$ is in $\ti V_2\cap \ti V_2^\perp= W_2\cap W_2^\perp$.  So $w_1\in W_1^\perp\cap W_2^\perp=0$, a contradiction. Hence $z_2=0$, i.e., $w_1= z_1\in \ti V_1$. Thus we have proved that $W_1^\perp\subset \ti V_1$. Since they both have dimension $1$, $W_1^\perp= \ti V_1$.  This proves the claim. 
 \end{proof}

\bthm {\bf [DT for the $\bn1$-flow with  ${\rm rank}(\pi)=1$]} \label{bea}\

\ni
Let $F(x, t, \cdot) \in (\bbn1)_+$ be a frame of a solution $q$ of the $(2j-1)$-th $\bn1$-flow \eqref{ah}, $\pi$ a rank one projection of $\R^{n+1,n}$ onto $V_1$ along $V_2$ satisfying \eqref{cua}, $\a_1\not=\a_2$ real constants, $g_{\a_1, \a_2, \pi}$  defined by \eqref{bd}, and $\tilde V_i(x,t)=F(x, t, \a_i)^{-1}V_i, i=1,2$. Assume there exists an open subset $\co$ of $\R^2$ containing $(0,0)$ such that  
\beq\label{iv}
\R^{n+1,n}=\ti V_1(x,t)\oplus \ti V_2(x,t)
\eeq 
for all $(x,t)\in \co$. Let $\ti \pi(x,t)$ be the projection onto $\ti V_1(x,t)$ along $\ti V_2(x,t)$. Then the projection   $\ti\pi(x,t)$  satisfies \eqref{cua},   
 \beq\label{ea}
 \ti{q}:= q+(\a_1-\a_2)[\b, \ti\pi-\ti\pi^\sharp]
 \eeq is a smooth solution of \eqref{ah} defined on $\co$, and
\beq\label{ec}
 \ti{F}(x, t, \l)=g_{\a_1, \a_2, \pi}(\l)F(x,t,\l)g^{-1}_{\a_1,\a_2,\ti\pi(x,t)}(\l)
 \eeq is a frame of the solution $\ti q$.
 \ethm

\begin{proof}
Fix $(x,t)\in \co_1$, we apply Theorem  \ref{ij} to $F(x,t,\cdot)$ and $g_{\a_1,\a_2,\pi}$. Then this theorem follows from Theorem \ref{gx}. 
\end{proof}

Theorem \ref{bea} can also be stated as follows:

\bthm  
Let $q$ be a solution of the $(2j-1)$-th $\bn1$-flow \eqref{ah}. Then we have:
\ben
\item The following system is solvable for $y:\R^2\to \R^{n+1, n}$ 
\beq
(DT)_{q, \a} \bca y_x= -A(q,\a) y,\\ y_t= -B(q,\a)y,\eca
\eeq
where  $\a\in \R$ is a constant parameter and 
 \beq\label{mn}
 A(q,\l)= J_B(\l) + q, \quad B(q,\l)= (T^{2j-1}(q,\l))_+.
 \eeq
 \item Let $\{v_1, \ldots, v_{2n+1}\}$ be a basis of $\R^{n+1,n}$ such that  $V_1=\R v_1$ and $V_2=\oplus_{i=2}^{2n+1} \R v_i$ satisfy \eqref{cu}. Given $\a_1\not=\a_2\in \R$, let $y_1$ be the solution of (DT)$_{q,\a_1}$ with $y_1(0)=v_1$, $y_i$ the solution of (DT)$_{q,\a_2}$ with $y_i(0)= v_i$ for $2\leq i\leq 2n+1$, and $\ti\pi(x,t)$ the projection onto $\R y_1(x,t)$ along $\oplus_{i=2}^{2n+1} \R y_i(x,t)$. Then 
$\ti q$ defined by \eqref{ea}
  is again a solution of \eqref{ah}.
  \een
\ethm

Recall that DT (BT resp.) for a soliton equation construct new solutions from a given solution $q$ of the soliton equation algebraically from $q$ and a solution of a linear system of PDE (a non-linear first order system resp.). 
Next we use DT to construct B\"acklund transformations for the $\bn1$-hiearchy. 
 
\bthm \label{ms} {\bf [B\"acklund transformation for the $\bn1$-hierarchy]}\

\ni Let $q$ be a solution of the $(2j-1)$-th $\bn1$-flow \eqref{ah}, and $\a_1\not=\a_2\in \R$.  Then we have the following:
\ben
\item The following
\beq\label{mm}
(BT)_{q,\a_1, \a_2} \bca 
\ti\pi_x= \ti \pi A(q,\a_2) - A(q+(\a_1-\a_2)[\b, \ti\pi-\ti\pi^\sharp], \a_2) \ti\pi,\\ 
\ti\pi_t= \ti\pi B(q,\a_2) -B(q+(\a_1-\a_2)[\b, \ti\pi-\ti\pi^\sharp], \a_2) \ti \pi,
\eca
\eeq
is a first order system for $\ti\pi:\R^2\to \Pi$ and is solvable, where 
$$\Pi= \{\pi\n \pi^2=\pi, {\rm rank}(\pi)=1, \pi\pi^\sharp= \pi^\sharp\pi=0\},$$  $A(q,\l)$ and $B(q,\l)$ are given by \eqref{mn}. 
 \item If $\ti\pi$ is a solution of \eqref{mm}, then 
$\ti q= q+(\a_1-\a_2)[\b, \ti\pi-\ti\pi^\sharp]$ is again a solution of \eqref{ah}. 
\een
\ethm

\begin{proof} Let $F$ be a frame of the solution $q$, and $\pi, \ti\pi, \ti q$ as in  Theorem \ref{bea}. Then 
 \beq\label{mp}
 \ti F(x,t,\l)= g_{\a_1,\a_2,\pi}(\l) F(x,t,\l) g_{\a_1, \a_2, \ti\pi(x,t)}^{-1}
 \eeq
 is a frame of the new solution $\ti q$ defined by \eqref{ea}.
Let
 \begin{align*}
 \Theta&= F^{-1}\rd F= A(q,\l) \rd x+ B(q,\l)\rd t,\\
  \ti \Theta&= \ti F^{-1}\rd \ti F= A(\ti q, \l) \rd x+ B(\ti q, \l)\rd t,
  \end{align*}
  where $\ti q= q+ (\a_1-\a_2)[\b, \ti\pi-\ti\pi^\sharp]$. 
 Set $\ti g=g_{\a_1,\a_2,\ti\pi}$. Use \eqref{mp} and a direct computation to see that  
\beq\label{mq}
 \ti \Theta = \ti g\Theta \ti g^{-1}- \rd\ti g \ti g^{-1}.
 \eeq
 So we have  
 \beq\label{mr}
 \rd\ti g= \ti g\Theta -\ti\Theta \ti g.
 \eeq
  By \eqref{bd}, we have
  \beq\label{mra}
  \ti g= g_{\a_1,\a_2,\ti\pi} = \I_{2n+1} +\frac{\a_2-\a_1}{\l-\a_2}\ti\pi +\frac{\a_1-\a_2}{\l-\a_1} \ti\pi^\sharp.
  \eeq
  Substitute \eqref{mra} into \eqref{mr} and compare the residues at $\l=\a_2$ of \eqref{mr} to see that $\ti\pi$ satisfies \eqref{mm}. 
  
  Next we prove that \eqref{mm} is a first order system in $\ti \pi$, i.e., the right hand side does not involve derivatives of $\ti\pi$.  First we claim that $(T^{2j-1}(\ti q,\l))_+$ is given algebraically by $(T^{2j-1}(q,\l))_+$ and $\ti\pi$. To see this, we use notation in Theorem \ref{zc}:
\begin{align*}
&M^{-1}(q,\l)(\p_x+J_B)M(q,\l)= \p_x+J_B+ q,\\
& T(q,\l)= M^{-1}(q,\l) J_B(\l) M(q,\l),\\
&M^{-1}(\ti q,\l)(\p_x+J_B)M(\ti q,\l)= \p_x+J_B+ \ti q,\\
& T(\ti q,\l)= M^{-1}(\ti q,\l) J_B(\l) M(\ti q,\l).
\end{align*}
Set $$M:=M(q,\l), \quad \ti M:= M(\ti q, \l), \quad \ti g= g_{\a_1,\a_2, \ti\pi(x,t)}(\l).$$ Since $\ti F$ is a frame for $\ti q$, 
\begin{align*}
\ti F^{-1}\ti F_x&= J_B+ \ti q\\
&= \ti g F^{-1}F_x \ti g^{-1}= \ti g (J_B+ q) \ti g^{-1} -\ti g_x \ti g^{-1}\\
&= \ti g(\p_x+ J_B+ q)\ti g^{-1} = \ti g M^{-1} (\p_x+J_B) ^{-1}M \ti g^{-1}.
\end{align*}
So we have $\ti M= M \ti g^{-1}$ and 
$$\ti T= \ti M^{-1} J_B \ti M= \ti g M^{-1} J_B M\ti g^{-1}= \ti g T \ti g^{-1}.$$
Hence $\ti T^{2j-1}= \ti g T^{2j-1} \ti g^{-1}$. This shows that $A(\ti q,\l)= T(\ti q,\l)$ and  $B(\ti q,\l)= (T^{2j-1}(\ti q,\l))_+$ ares given by algebraic formulas in terms of $(T^{2j-1}(q,\l))_+$ and $\ti \pi$. In particular, we have proved that $B(\ti q, \a_1)$ depends on $\ti\pi$ algebraically (no derivatives of $\ti \pi$ are involved). Thus system \eqref{mm} is a first order non-linear system in $\ti\pi$. 
  
  Since $\ti\pi(0)$ can be any element in $\Pi$, the first order system \eqref{mm} is solvable. This proves (1).
  
  Conversely, let $\hat \pi$ be the solution of \eqref{mm} with $\hat\pi(0)=\pi_0\in \Pi$. Let $\ti\pi$ be constructed from $\pi_0$ as in Theorem \ref{bea}. We have proved above that $\ti\pi$ also satisfies \eqref{mm}. Since they have the same initial data, by the uniqueness part of Frobenius Theorem we conclude that $\ti\pi= \hat \pi$. This proves (2). 
   \end{proof}

\bthm {\bf [DT for the $\bn1$-KdV flows]}\ 

\ni
Let $E(x,t,\l)$ be a frame of a solution $u$ of the $(2j-1)$-th $\bn1$-KdV flow \eqref{ai}, $\pi$ the projection onto $V_1$ along $V_2$ satisfying \eqref{cua}, $\dim(V_1)=1$, and $\a_1\not= \a_2$ real numbers. Let 
$$\ti V_i(x,t)= E(x,t,\a_i)^{-1}(V_i), \quad i=1,2,$$
and $\ti\pi(x,t)$ the projection onto $\ti V_1(x,t)$ along $\ti V_2(x,t)$. Then
\ben
\item $\hat E(x,t,\l):= g_{\a_1,\a_2,\pi}(\l) E(x,t,\l) g_{\a_1, \a_2, \ti\pi(x,t)}^{-1}(\l)$ is holomorphic,
\item $\hat E^{-1}\hat E_x= J_B+ \hat q$ for some $q(x,t)\in \cb_n^+$,
\item there exists a unique $\hat\D(x,t)$ such that $\ti u:=\hat\D\ast \hat q$ is $V_n$-valued,
\item $\ti u$ is a solution of the $(2j-1)$-th $\bn1$-KdV flow \eqref{ai},
\item $\ti E(x,t,\l)= \hat E(x,t,\l)\hat \D(x,t)^{-1}$ is a frame of $\ti u$.
\een
\ethm

\begin{proof} 
By Proposition \ref{do}, there exists  $\D:\R^2\to N_n^+$ such that $\D^{-1}\ast u= q$ is a solution of the $(2j-1)$-th $\bn1$-flow \eqref{ah} and $F= E\D$ is a frame of $q$. It follows from Theorem \ref{bea} that 
$$\ti F= g_{\a_1,\a_2,\pi} F g_{\a_1,\a_2, \ti\pi(x,t)}^{-1}$$
is a frame of a new solution $\ti q= q+(\a_1-\a_2)[\b, \ti\pi_1-\ti\pi_1^\sharp]$ of \eqref{ah}, where $\ti\pi_1(x,t)$ is the projection onto $\hat V_1(x,t)$ along $\hat V_2(x,t)$ and 
$$\hat V_i(x,t)= F(x,t,\a_i)^{-1}(V_i), \quad i=1,2.$$
 By Proposition \ref{cga}, there exists a $\ti \D:\R^2\to N_n^+$ such that $\ti u:= \ti \D\ast \ti q$ is a solution of \eqref{ai} with frame $\ti E= \ti F \ti \D^{-1}$. 
Set $g= g_{\a_1,\a_2,\pi}$ and $\ti g_1(x,t,\l)= g_{\a_1,\a_2, \ti\pi_1(x,t)}(\l)$. 
Then we have
$$gE= gF\D^{-1}= \ti F \ti g_1 \D^{-1}= \ti  E\ti \D\ti g_1\D^{-1}= (\ti E \ti \D \D^{-1}) (\D \ti g_1\D^{-1}).$$
Note that
$\hat E:=\ti E \ti \D \D^{-1}\in (\bbn1)_+$ and 
$\hat g:= \D \ti g_1 \D^{-1}\in \R\bbn1$. 
  This proves the theorem.
\end{proof}

If the rank of $\pi$ is bigger than $1$, the proof of Theorem \ref{ij} also gives a DT of \eqref{ah}. But the formula of the new solutions is a bit more complicated:

\bthm{\bf [DT for the the $\bn1$-flow with rank$(\pi)>1$]}\label{ed}\

\ni
Let $F, q, \a_1, \a_2$ be as in Theorem \ref{bea}, and $\pi$ a projection of $\R^{n+1,n}$ onto $V_1$ along $V_2$ satisfying \eqref{cua}. Let
$$\ti V_i(x,t)= F(x,t,\a_i)^{-1}(V_i), \quad i=1, 2.$$ Assume that  there exists an open subset $\co$ of $\R^2$ such that 
$\ti V_1(x,t)\cap \ti V_2(x,t)=\{0\}$. Let
 $\ti\pi(x,t)$ be the projection onto $\ti V_1(x,t)$ along $\ti V_2(x,t)$,  
\begin{align*}
k(x,t,\l)&= h_{\a_1,\a_2,\pi}(\l) F(x,t,\l) h^{-1}_{\a_1, \a_2, \ti\pi(x,t)}(\l),\\
W_i(x,t)& = k(x,t,\a_i)^{-1}(V_i^\perp), \quad i=1,2.
\end{align*} 
Assume that $W_1(x,t)\cap W_2(x,t)=\{0\}$ for all $(x,t)\in \co_1\subset \co$. Let $\o(x,t)$ be the projection onto $W_2(x,t)$ along $W_1(x,t)$. Then $\ti q= q + (\a_1-\a_2)[\b, \ti\pi-\o]$ is a solution of \eqref{ah} defined on $\co_1$ and 
$$\ti F(x,t,\l) := g_{\a_1, \a_2,\pi} F(x,t,\l) h^{-1}_{\a_1,\a_2,\ti\pi(x,t)}(\l) h^{-1}_{\a_2,\a_1,\o(x,t)}(\l)$$ 
is a frame of $\ti q$.
\ethm

 Since we need to use Corollary \ref{ijb} to compute $W_i(x,t)$ given in Theorem \eqref{ed}, the DT formula is more complicated than the case when the rank of $\pi$ is one. 

\section{Permutability Formula}\label{ipa}
 
 In this section, we give a Permutability formula for Darboux transforms of the $\bn1$-flows. 

 First we write down some relations among simple elements.

\blem\label{bm} 
Let $\a_1,\b_1, \a_2,\b_2$ be distinct real constants, $\pi_1, \pi_2$ rank one projections of $\R^{n+1, n}$ satisfying \eqref{cua}, and $\tau_1,\tau_2$ projections defined by
\beq\label{gm}
\bca \Im(\tau_1)=g_{\a_2, \b_2, \pi_2}(\a_1)\Im(\pi_1), \quad
\Ker(\tau_1)=g_{\a_2, \b_2, \pi_2}(\b_1)\Ker(\pi_1), \\
 \Im(\tau_2)=g_{\a_1, \b_1, \pi_1}(\a_2)\Im(\pi_2), \quad
\Ker(\tau_2)=g_{\a_1, \b_1, \pi_1}(\b_2)\Ker(\pi_2).
\eca
\eeq
Then $\tau_1, \tau_2$ satisfy \eqref{cua} and 
\beq\label{gf}
g_{\a_2, \b_2, \tau_2}g_{\a_1, \b_1, \pi_1}=g_{\a_1, \b_1, \tau_1}g_{\a_2, \b_2, \pi_2}.
\eeq
Conversely, if $\pi_i$ and $\tau_i$ satisfy \eqref{cua} and \eqref{gf} then \eqref{gm} is true.
\elem

\begin{proof} 
Apply Theorem \ref{ij} to $g_{\a_2,\b_2,\pi_2}$ and $f=g_{\a_1, \b_1,\pi_1}^{-1}$ and use \eqref{bda} to see that 
$g_{\a_2,\b_2,\pi_2} g_{\a_1, \b_1, \pi_1}^{-1} = \ti f g_{\a_2, \b_2, \tau_2}$. 
Hence we have $g_{\a_1,\b_1, \pi_1} g_{\a_2, \b_2, \pi_2}^{-1}= g_{\a_2, \b_2, \tau_2} ^{-1}\ti f^{-1}$.
It follows from Theorem \ref{ij} again that $\ti f^{-1}= g_{\a_1, \b_1,\tau_1}$. This proves that 
$g_{\a_2,\b_2,\pi_2} g_{\a_1, \b_1, \pi_1}^{-1} = g_{\a_1, \b_1, \tau_1}^{-1} g_{\a_2, \b_2, \tau_2}$, which is \eqref{gf}.
The converse follows from Theorem \ref{ij}.  
\end{proof}

Note that if $\pi$ is a projection onto $V_1$ along $V_2$, $\{v_1, \ldots, v_k\}$ is a basis of $V_1$, and $\{v_{k+1}, \ldots, v_{2n+1}\}$ is a basis of $V_2$, then 
\beq\label{mh}
\pi= (v_1, \ldots, v_k, 0, \ldots, 0)(v_1, \ldots, v_{2n+1})^{-1}.
\eeq
 So it follows from \eqref{gm} that $\tau_i$ can be obtained algebraically from $\pi_1$ and $\pi_2$.

\bthm\label{ev} {\bf [Permutability for DT of the $\bn1$-flows]} \ 

\ni Let $\a_i,\b_i, \pi_i, \tau_i$ be as in Lemma \ref{bm}, $q$ a solution of the $(2j-1)$-th $\bn1$-flow \eqref{ah}. Then we have the following:
\ben
\item $g_{\a_2, \b_2, \tau_2}\bu(g_{\a_1, \b_1, \pi_1}\bu q)= g_{\a_1, \b_1,\tau_1}\bu(g_{\a_2,\b_2,\pi_2}\bu q)$.
\item Let $F(x,t,\l)$ be the frame of the solution $q$ with $F(0,0, \l)=\I_{2n+1}$, and $\ti\pi_i(x,t) (i=1, 2)$ the projection onto $F(x,t, \a_i)^{-1}(\Im\pi_i)$ along $F(x,t,\b_i)^{-1}(\Ker\pi_i)$. Then 
\begin{align*}
q_1&:=g_{\a_1, \b_1, \pi_1} \bu q = q+ (\a_1-\b_1)[\b, \ti \pi_1-\ti\pi_1^\sharp], \\
q_2 &:=g_{\a_2, \b_2, \pi_2} \bu q = q+ (\a_2-\b_2)[\b, \ti \pi_2-\ti\pi_2^\sharp].
\end{align*}
\item Let $\ti\tau_1(x,t), \ti\tau_2(x,t)$ be the projections defined by 
\begin{align}
&\Im(\ti\tau_1)=g_{\a_2, \b_2, \ti\pi_2}(\a_1)\Im(\ti \pi_1), \quad
\Ker(\ti\tau_1)=g_{\a_2, \b_2, \ti\pi_2}(\b_1)\Ker(\ti \pi_1), \label{gp}\\
&
\Im(\ti\tau_2)=g_{\a_1, \b_1, \ti \pi_1}(\a_2)\Im(\ti \pi_2), \quad
\Ker(\ti\tau_2)=g_{\a_1, \b_1, \ti \pi_1}(\b_2)\Ker(\ti \pi_2).  \label{gq}
\end{align} 
Then we have
\begin{align*}
q_{12}&:=g_{\a_2, \b_2, \tau_2}\bu(g_{\a_1, \b_1, \pi_1}\bu q)= g_{\a_1, \b_1, \tau_1}\bu (g_{\a_2, \b_2, \pi_2}\bu q)\\
&= q_1+ (\a_2-\b_2)[\b, \ti \tau_2-\ti \tau_2^\sharp]= q_2+(\a_1-\b_1)[\b, \ti\tau_1-\ti\tau_1^\sharp].
\end{align*}
\een
In particular, $q_{12}$ can be obtained algebraically from $\ti\pi_1$ and $\ti\pi_2$. 
\ethm

\begin{proof} Let $F$ be the frame of the solution $q$ of \eqref{ah} with $F(0,0,\l)=\I_{2n+1}$. Theorem \ref{bea} implies that 
\begin{align*}
F_1& = g_{\a_1, \b_1, \pi_1}F g_{\a_1, \b_1, \ti\pi_1}^{-1},\\
F_2& = g_{\a_2, \b_2, \pi_2}F g_{\a_2, \b_2, \ti\pi_2}^{-1},
\end{align*}
are the frames of $q_1$ and $q_2$ respectively.  Apply Theorem \ref{bea} to $q_2$ and $q_1$ to see that there are projections $\hat\tau_1(x,t)$ and $\hat\tau_2(x,t)$ such that
\begin{align*}
F_{12} &=g_{\a_2, \b_2, \tau_2}g_{\a_1, \b_1, \pi_1} F g^{-1}_{\a_1, \b_1, \hat \pi_1}g^{-1}_{\a_2, \b_2, \hat\tau_2},\\
F_{21} & = g_{\a_1, \b_1, \tau_1} g_{\a_2, \b_2, \pi_2} Fg^{-1}_{\a_2, \b_2, \ti\pi_2} g^{-1}_{\a_1, \b_1, \hat\tau_1}
\end{align*}
are the frames of $q_{12}$ and $q_{21}$ respectively.  Let $f=g_{\a_2, \b_2, \tau_2}g_{\a_1, \b_1, \pi_1}$. By assumption, 
$f=  g_{\a_1, \b_1, \tau_1} g_{\a_2, \b_2, \pi_2}$. So we obtain 
$$fF= F_{12} g_{\a_2, \b_2, \hat\tau_2}g_{\a_1, \b_1, \ti\pi_1} = F_{21}g_{\a_1, \b_1, \hat\tau_1} g_{\a_2,\b_2, \ti \pi_2}.$$
This gives two factorizations of $fF$ as the product of elements in $(\bbn1)_+$ and $\R\bbn1$. It follows from Proposition \ref{ho} that we have $F_{12}= F_{21}$ and
\beq\label{go} g_{\a_2,\b_2, \hat\tau_2} g_{\a_1, \b_1, \ti\pi_1} = g_{\a_1, \b_1, \hat\tau_1} g_{\a_2,\b_2, \ti \pi_2}. 
\eeq
It follows from \eqref{go} and Lemma \ref{bm}(2) that $\hat\tau_i=\ti\tau_i$ defined by \eqref{gp} and \eqref{gq}.  Since $F_{12}=F_{21}$, $q_{12}= q_{21}$. Formulas for $q_1, q_2, q_{12}$  follow from Theorem \ref{bea}. 
\end{proof}

\bs
\section{Scaling Transforms}\label{mt}

In this section, we construct scaling transforms and give relations between DTs and scaling transforms for the $\bn1$-flows. 

\bthm\label{qi} Let $F(x, t, \l)$ be the frame of the solution $q$ of the $(2j-1)$-th $\bn1$-flow with $F(0, 0, \l)=I_{2n+1}$. Let $r \in \R \bh \{0\}$ and 
\beq\label{cb}\G(r)=\diag(1, r, \ldots, r^{2n}).
\eeq
 Then
\ben
\item $(r \odot q) (x, t) :=r\G(r)^{-1}q(rx, r^{2j-1}t)\G(r)$ is a solution of the $(2j-1)$-th $\bn1$-flow.
\item $(r \odot F)(x, t, \l):= \G(r)^{-1}F(rx, r^{2j-1}t, r^{-2n}\l)\G(r)$ is the frame of $r \odot q$ such that $\hat F(0, 0, \l)=\I_{2n+1}$.
\een
\ethm

\begin{proof}  A direct computation implies that 
\begin{align}
&\G(r)\rho_n\G(r)= r^{2n+1} \rho_n, \label{cb1}\\
& r\G(r)^{-1} J_B(r^{-2n}\l) \G(r)= J_B(\l). \label{ca3}
\end{align}
  Let $M(q,\l)$ be a solution of \eqref{bb} as in Theorem \ref{zc}, and 
$$\ti M: = \G(r)^{-1}M(q, r^{-2n}\l)(rx) \G(r).$$ Since $M$ satisfies $M^t\rho_n M= \rho_n$,
use \eqref{cb1} to compute directly to see that $\ti M^t\rho_n \ti M= \rho_n$. It is clear that $\overline{\ti M(\cdot, \bar\l)}= \ti M(\cdot, \l)$. 
 Compute directly to get  
$$\ti M^{-1} (\p_x + J_B(\l)) \ti M = \p_x+ J_B(\l) + r\odot q.$$
So $\ti M$ satisfies all conditions of Theorem \ref{zc} (1), i.e., $\ti M(x,\l)= M(r\odot q, \l)$. By Theorem \ref{zc}(2),  we have 
\begin{align}
&T(r\odot q, \l)(x)= \G(r)^{-1} T(q, r^{-2n}\l) (rx) \G(r), \label{ca1}\\
& T^{2j-1}(r\odot q, \l)(x)= \G(r)^{-1} T^{2j-1}(q, r^{-2n}\l) (rx) \G(r),\label{ca2}
\end{align}
  Let $\hat F(x,t,\l)= \G(r)^{-1}F(rx, r^{2j-1}t, r^{-2n}\l)\G(r)$. 
 Use \eqref{cb1}, \eqref{ca3}, \eqref{ca1}, \eqref{ca2} and a direct computation to see that 
 $\hat F$ satisfies
 $$\bca\hat F_x= \hat F(J_B+ r\odot q),\\
 \hat F_t= \hat F(P^{2j-1}(r\odot q,  r^{-2n}\l))_+. 
\eca
$$
Hence $r\odot q$ is a solution of the $(2j-1)$-th $\bn1$-flow and $\hat F$ is the frame of $r\odot q$ with $\hat F(0,0,\l)=\I_{2n+1}$.  
\end{proof}

\bcor \label{qb} Let $u=\sum_{i=1}^n u_i \b_i$ be a solution of the $(2j-1)$-th $\bn1$-KdV flow, where $\b_i$ be as in \eqref{cj}. Let $r \in \R \backslash \{0\}$, $\G(r)=\diag(1, r, \ldots, r^{2n})$ and 
\beq\label{qe}
(r \cdot u_i)(x, t)=r^{2i}u_i(rx, r^{2j-1}t), \quad 1 \leq i \leq n.
\eeq
Then
\ben 
\item  $r \odot u=\sum_{i=1}^n (r \cdot u_i)\b_i$ is a solution of the $(2j-1)$-th $\bn1$-KdV flow,
\item if $E(x, t, \l)$ is a frame of $u$, then
$$(r \odot E)(x, t, \l):=\G(r)^{-1}E(rx, r^{2j-1}t, r^{-2n}\l)\G(r)$$
is a frame of $r \odot u$.
\item $r\odot u$ defines an action of the multiplicative group $\R^+$ on the space of solutions of the $(2j-1)$-th $\bn1$-KdV flow.
\een
\ecor

The next Theorem gives a relation between the scaling transforms and Darboux transforms for the $\bn1$-flows. 

\bthm  Let $F(x, t, \l)$ be the frame of the solution $q$ of the $(2j-1)$-th $\bn1$-flow with $F(0, 0, \l)=I_{2n+1}$. Let $r \in \R \bh \{0\}$ and $\a \in \R$, $\pi$ a rank one projection onto $V_1$ along $V_2$ satisfying \eqref{cua}, $\hat V_i=\G(r)V_i$ for $i=1,2$, and $\hat \pi$ the projection onto $\hat V_2$ along $\hat V_1$.  
 Then $\hat \pi$ satisfies \eqref{cua} and 
 $$g_{r^{-2n}, r^{-2n}\a, \hat \pi}\bu q(x, t)=r^{-1} \odot (g_{1, \a, \pi} \bu r \odot q(x, t)).$$
\ethm

\begin{proof} Note that $\hat \pi= \G(r)\pi \G(r)^{-1}$ and $\li \G(r) x, \G(r)y \ri=r^{2n} \li x, y\ri $ for any $x, y \in \R^{n+1, n}$. Hence  $\hat V_1$ and $\hat V_2$ satisfies \eqref{cu}. It follows from Lemma \ref{bc} that $\hat \pi$ satisfies \eqref{cua}.   

It follows from a direction computation that we have 
\begin{align*}
&\G(r) \rho_n=r^{2n}\rho_n \G(r)^{-1},\\
&\hat \pi^\sharp=\rho_n\hat \pi^t \rho_n=\rho_n \G(r)^{-1}\pi^t \G(r)\rho_n=\G(r)\rho_n\pi^t \rho_n \G(r)^{-1}=\G(r)\pi^{\sharp}\G(r)^{-1}.
\end{align*}

By Theorem \ref{qi}, 
\beq\label{zab}
\ti F(x, t, \l)=\G(r)^{-1}F(rx, r^{2j-1}t, r^{-2n}\l)\G(r)
\eeq
is the frame for $\ti q(x, t)= r \odot q(x, t)$ satisfying $\ti F(0, 0, \l)=I_{2n+1}$. 

Set 
\beq\label{zaa}
\ti V_1(x, t)= \ti F(x, t, 1)^{-1}V_1, \quad \ti V_2(x, t)= \ti F(x, t, \a)^{-1}V_2.
\eeq
It follows from Theorem \ref{bea} that
$$F_1(x, t, \l)=g_{1, \a, \pi}(\l) \ti F(x, t, \l)g_{1, 0, \ti \pi(x,t)}^{-1}(\l)$$
is a frame for $q_1(x, t):= g_{1, \a, \pi}\bu \ti q(x, t)$, 
where $\ti \pi(x,t)$ is the projection  of $\R^{n+1, n}$ onto $\ti V_1(x,t)$ along $\ti V_2(x,t)$. 

By Theorem \ref{qi}, we have 
$$F_2(x, t, \l)=\G(r)F_1(r^{-1}x, r^{-(2j-1)}t, r^{2n}\l)\G(r)^{-1}$$
is a frame for $q_2(x, t):=r^{-1} \odot q_1(x, t)$.

A direct computation implies that 
\begin{align*}
&\quad  F_2(x, t, \l) = \G(r)F_1(r^{-1}x, r^{-(2j-1)}t, r^{2n}\l)\G(r)^{-1} \\
 &  = \G(r)(g_{1, \a, \pi}(r^{2n}\l) \ti F(r^{-1}x, r^{-(2j-1)}t, r^{2n}\l)g^{-1}_{1, \a, \ti \pi(r^{-1}x, r^{-(2j-1)t})}(r^{2n}\l)\G(r)^{-1} \\
& = \G(r)(g_{1, \a, \pi}(r^{2n}\l)  \G(r)^{-1} F(x, t, \l)\G(r)g^{-1}_{1, 0, \ti \pi(r^{-1}x, r^{-(2j-1)t})}(r^{2n}\l)\G(r)^{-1} \\
& = g_{r^{-2n}, r^{-2n}\a, \hat \pi}F(x, t, \l)g^{-1}_{r^{-2n}, r^{-2n}\a, \hat \theta},
\end{align*}
where $\hat \theta$ is the projection of $\R^{n+1, n}$ onto $ W_1=\G(r)\ti V_1(r^{-1}x, r^{-(2j-1)}t)$ along 
$W_2=\G(r)\ti V_2(r^{-1}x, r^{-(2j-1)}t)$. By \eqref{zab} and \eqref{zaa}, we have
\begin{align*}
\G(r)\ti V_1(r^{-1}x, r^{-(2j-1)}t) & = \G(r)\ti F^{-1}(r^{-1}x, r^{-(2j-1)}t, 1)V_1 \\
&= F^{-1}(x, t, r^{-2n})\G(r)V_1 \\
& =F^{-1}(x, t, r^{-2n})\hat V_1
\end{align*}
A similar computation shows that $W_2=F^{-1}(x, t, r^{-2n}\a)\hat V_2$. 

It follows from Theorem \ref{bea} that $F_2(x, t, \l)$ is a frame for $g_{r^{-2n}, r^{-2n}\a, \hat \pi} \bu q(x, t)$. This proves the theorem.
\end{proof}

  \bs
  \section{Explicit soliton solutions} \label{mu}

We apply Darboux transforms for the $\bn1$-flows to the vacuum solution to construct explicit soliton solutions for the $\bn1$-, $\bn1$-KdV, and isotropic curve flows of B-type. 

First we review the relation between isotropic curve flows of type B and $\bn1$-flows in \cite{TWd}. Let 
$$\cm_{n+1,n}(S^1)=\{\g\in \cm_{n+1,n}\,\n\, \g(x+2\pi)= \g(x)\,\, \forall\, x\in \R\},$$
where $\cm_{n+1.n}$ is defined by \eqref{jc}.
\ben
\item There is a natural Poisson structure on $\cm_{n+1,n}$ so that the Hamiltonian equation for the functional $\hat F:\cm_{n+1,n}(S^1)\to \R$ defined by $\hat F(\g)= F(u)$ is 
$$\g_t= gP_u(\K F (u))e_1,$$
where $g$ and $u$ are the isotropic moving frame and curvature. 
\item Let $F_{2j-1}(u)=-\oint \tr(T_{2j-1,-1}(u)\b)\rd x$. Then 
$$\K F_{2j-1}(u)= \pi_0(T_{2j-1,0}(u)),$$ where $\pi_o$ is the natural projection of $o(n+1,n)$ to $V_n$. So the Hamiltonian flow for $\hat F_{2j-1}$ on $\cm_{n+1,n}(S^1)$ is 
$$\g_t= gP_u(\pi_0(T_{2j-1,0}(u))) e_1= g T_{2j-1,0}(u) e_1.$$
\item In particular, the third isotropic curve flow \eqref{ma} on $\cm_{2,1}$ and \eqref{jb} on $\cm_{3,2}$ are the Hamiltonian flows for $F(\g)=-\oint q^2 \rd x$ on $\cm_{2,1}$ and $H(\g)= \oint \frac{3}{2} u_1^2 + 2 u_2\rd x$ on $\cm_{3,2}$ respectively.
   \een
   
   \bdefn
The $(2j-1)$-th isotropic curve flow of type B is the following flow on $\cm_{n+1,n}$
\beq\label{kh}
\g_t= g T_{2j-1,0}(u) e_1,
\eeq
where $T_{2j-1,0}(u)$ is defined by \eqref{dg}, $g(\cdot, t)$ and $u(\cdot, t)$ are the isotropic moving frame and curvature along $\g(\cdot, t)$, and 
$$e_1=(1,0,\ldots, 0)^t\in \R^{n+1,n}.$$
\edefn

\bthm\label{kj}  (\cite{TWd})\ 

\ben
\item If $\g$ is a solution of the $(2j-1)$-th isotropic curve flow \eqref{kh}, then its isotropic curvature $u$ is a solution of the $(2j-1)$-th $\bn1$-KdV flow \eqref{ai}.
\item If $F(x,t,\l)$ is a frame of a solution $q$ of the $(2j-1)$-th $\bn1$-flow \eqref{ah}, then $\g(x,t):= F(x,t,0)e_1$ is a solution of \eqref{kh}. Moreover, the isotropic curvature $u(\cdot, t)$ of $\g(\cdot, t)$ is a solution of the $(2j-1)$-th $\bn1$-KdV flow \eqref{ai}.
\een
\ethm

\bthm {\bf [DT for isotropic curve flow of B type]} \label{mw}\

\ni Let $\g$ be a solution of the $(2j-1)$-th isotropic curve flow \eqref{kh} on $\cm_{n+1,n}$, and $g(\cdot, t)$ and $u(\cdot, t)$ the isotropic moving frame and curvature along $\g(\cdot, t)$. Let $E(x,t,\l)$ be the frame of the solution $u$ of the $(2j-1)$-th $\bn1$-flow \eqref{ah} with $E(0,0,\l)= g(0,0)$,  $\D:\R^2\to N_n^+$  with $\D(0,0)=\I_{2n+1}$, and $F= E\D$ the frame of the solution $q=\D^{-1}\ast q$ of the $(2j-1)$-th flow \eqref{ah} as in Proposition \ref{do}. 
Let $\a_1\not= \a_2\in \R$ be constants, $\pi$ a projection of $\R^{n+1,n}$ satisfying \eqref{cua}, and $\ti\pi(x,t)$ the projection onto $\ti V_1(x,t)= F(x,t,\a_1)^{-1}(\Im\pi)$ along $\ti V_2(x,t)= F(x,t,\a_2)^{-1}(\Ker\pi)$. 
\ben 
\item If $\a_1\a_2\not=0$, then 
$$\ti \g_{\a_1,\a_2}= g\D (\I_{2n+1} -\frac{\a_2-\a_1}{\a_2} \ti\pi^\sharp -\frac{\a_1-\a_2}{\a_1} \ti \pi)e_1$$
is a solution of \eqref{kh}. 
\item If $\a_2=0$, then $\ti \g_{\a,0}= Ye_1$ is is a solution of \eqref{kh}, where 
\begin{align*}
Y:&=(\pi g \ti\pi+(\pi^\sharp)^\perp g \ti\pi^\perp+\pi^\sharp g  \ti\pi^\sharp)\\
&-\a(\pi Z_1 \ti\pi^\perp +(\pi^\sharp)^\perp Z_1 \ti\pi^\sharp) +\a^2\pi Z_2 \ti\pi^\sharp,\\
&Z_1(x,t)=\frac{\p F}{\p\l} (x,t,0), \quad Z_2(x,t)= \frac{1}{2} \frac{\p^2 F}{\p \l^2}(x,t,0).
 \end{align*}
 \een
\ethm

\begin{proof}
If $f\in (\bbn1)_+$ and $\ti F$ is a frame of the solution $\ti q$ of \eqref{ah}, then $f(\l) \ti F(x,t,\l)$ is a frame of $\ti q$. So (1) follows from Theorems \ref{bea} and \ref{kj}. 

 Theorem \ref{bea} implies that $\ti q=q+\a^2[\b, \ti\pi-\ti\pi^\sharp]$ is a solution of \eqref{ah} and 
 $$\ti F(x,t,\l) = g_{\a^2,0,\pi} F(x,t\l) g^{-1}_{\a^2,0, \ti\pi(x,t)}(\l)$$
 is a frame of $\ti q$. 
 Since $\ti F(x,t,\l)$ is holomorphic at $\l=0$, $\ti F(x,t,0)$ can be obtained by expanding $g_{\a^2,0,\pi}(\l)$, $F(x,t,\l)$, and $g_{\a^2,0,\ti\pi(x,t)}(\l)$ as a power series of $\l$ and compute the constant term.  A direct computation implies that $\ti F(x, t, 0)= Y(x,t)$.  Statement (2) follows from Theorem \ref{kj}.
\end{proof}

We compute $g_{\a_1,\a_2,\pi}\bu 0$ for the third $B_1^{(1)}$-KdV flow (the KdV flow) in the next two examples and obtain $1$-soliton solutions if $\a_2=0$ and $2$-soliton solutions if $\a_1>\a_2>0$. 

 \beg\label{hp} {\bf  [$g_{\a^2,0}\bu 0$ for $n=1$]}\ 

For $n=1$, we have $J_B^3(\l)=\l J_B(\l)$. 
The frame of the vacuum solution $q=0$ of the third $\hat B_1^{(1)}$-flow with $F(0,0,\l)=\I_3$ is 
 \begin{align*}
 F(x,t,\l)&= \exp(J_B(\l)x + \l J_B(\l)t) =F(x, t,z^2)\\
 &= \bpm  \frac{1}{2}(c_z+1) & \frac{z}{2}s_z & \frac{z^2}{4}(c_z-1) \\
\frac{1}{z} s_z &c_z &  \frac{z}{2}s_z \\
\frac{1}{z^2}(c_z-1) & \frac{1}{z} s_z & \frac{1}{2}(c_z+1) \epm,
\end{align*}
where $\l=z^2$, and
\beq\label{cv}
c_z(x, t)=\cosh(z x+z^3 t), \quad s_z(x, t)=\sinh (z x+z^3 t).
\eeq

Let $\{e_1, e_2, e_3\}$ be the standard basis on $\R^{3}$, and $\pi$ the projection onto $\R e_1$ along $\R e_2\oplus \R e_3$. Then $\pi$ satisfies \eqref{cua}.  Let $\a$ be nonzero real constant.  We apply Theorem \ref{bea} (DT of the $\bn1$-flow)  to the trivial solution $q=0$ of third $\hat B_1^{(1)}$-flow with frame $F(x,t,\l)$ and $g_{\a^2, 0,\pi}$.
 Let 
$$\ti p_1(x,t)= F(x,t,\a^2)^{-1}(e_1), \quad \ti p_i(x,t)= F(x,t,0)^{-1}(e_i), \, i=2,3,$$
and $\ti\pi(x,t)$ the projection onto $\R \ti p_1(x,t)$ along $\R \ti p_2(x,t) \oplus \R \ti p_3(x,t)$. So 
\beq\label{hp1} \ti p_1(x, t)=\bpm \frac{1}{2}(c_\a+1) \\ -\frac{1}{\a}s_\a \\ \frac{1}{\a^2}(c_\a-1)\epm , \quad \ti p_2(x, t)=\bpm 0\\ 1 \\ -x \epm , \quad \ti p_3(x, t)=e_3. 
\eeq 
So 
$$ \ti \pi = (\ti p_1, 0, 0)(\ti p_1, \ti p_2, \ti p_3)^{-1}=\bpm 1 & 0 & 0 \\ -\frac{2s_\a}{\a(c_\a+1)} & 0 & 0 \\ \frac{2(c_\a-1)}{\a^2(c_\a+1)} & 0 & 0 \epm. $$
By Theorem \ref{bea}, we obtain a new solution  
$$\ti q_\a:= g_{\a^2, 0, \pi}\bu 0 = (q_1)_{\a}(e_{11}-e_{33}) +(q_2)_{\a}(e_{12}+ e_{23})$$  of the third $\hat B_1^{(1)}$ flow, where 
\begin{align*}
(q_1)_{\a}=-\frac{\a s_\a}{c_\a+1}, \quad (q_2)_\a=-\frac{\a^2}{2}.
\end{align*}
and $c_a, s_\a$ are functions defined by \eqref{cv}. 
Moreover,  
$$\ti F(x,t,\l)= g_{\a^2, 0, \pi}(\l) F(x,t,\l) g_{\a^2, 0, \ti\pi(x,t)}(\l)^{-1}$$ is a frame of the solution $\ti q_\a$ of the third $\hat B_1^{(1)}$-flow.

Recall that the third $\hat B_1^{(1)}$-KdV is the KdV. 
By Corollary \ref{at},
\beq\label{bl}
y_{\a}=(q_2)_{\a}+\frac{1}{2}(q_1)_\a^2+\p_x(q_1)_{\a}=-\a^2\sech^2(\frac{\a}{2}x+\frac{\a^3}{2}t)
\eeq
 is a solution of the KdV \eqref{kdv}. 
 Note that these are the well-known 1-soliton solutions. 
 
 Next we write down 1-soliton solutions of the third isotropic curve flow \eqref{ma} on $\cm_{2,1}$ of type B. It follows from Theorem  \ref{kj}  that $F(x,t,0)e_1= e^{bx} e_1= (1, x, \frac{x^2}{2})^t$ with isotropic curvature $q\equiv 0$ of \eqref{ma}.  Use Theorem \ref{mw}(2) to compute directly to see that 
 \begin{align*}
\ti \g_{\a^2,0}(x,t):&=\left(g_{\a^2, 0, \pi}F(x,t, \l)g^{-1}_{\a^2, 0, \ti\pi(x,t)}\right)_{\l=0}e_1\\
&= \left(\pi F(x, t, 0)\ti \pi+(\pi^\sharp)^{\perp}F(x, t, 0)\ti \pi^{\perp}+\pi^\sharp F(x, t, 0)\ti \pi^{\sharp}\right)e_1 \\
& \quad -\a^2\left(\pi F_\l(x, t, 0)\ti \pi^{\perp}+(\pi^\sharp)^\perp F_\l(x, t, 0)\ti \pi^{\sharp}\right)e_1 \\
& \quad +\frac{1}{2}\a^4\pi \frac{\p ^2 F}{\p \l^2}(x, t, 0)\ti \pi^{\sharp}e_1\end{align*}
is a solution of the third isotropic curve flow \eqref{ma} on $\cm_{2,1}$. 
Note that 
 \begin{align*}
& F(x, t, 0)=e^{bx},  \quad F_\l(x, t, 0)=e^{bx}(\b x+bt), \\
& \frac{\p ^2 F}{\p \l^2}(x, t, 0)=e^{bx}((\b x+b t)^2+2\b t).
\end{align*}
Use the formula \eqref{hp1} and a direction computation implies that
\begin{align*}
\ti \g_{\a^2,0}(x, t) =  \bpm 1-\frac{\a s_{\a}(x,t)}{c_{\a}(x,t)+1}x+\frac{\a^2(c_{\a}(x,t)-1)}{4(c_{\a}(x,t)+1)}x^2  \\
\frac{2s_{\a}(x,t)}{\a(c_{\a}(x,t)+1)}-\frac{c_{\a}(x,t)-1}{c_{\a}(x,t)+1}x \\
\frac{2(c_{\a}(x,t)-1)}{\a^2(c_{\a}(x,t)+1)}
\epm
\end{align*}
is a solution of the third isotropic curve flow \eqref{ma}, whose isotropic curvature is the 1-soliton solution \eqref{bl} of the KdV, where $c_\a(x,t)$ and $s_\a(x,t)$ are defined by \eqref{cv}.   
\eeg

\beg\label{hp2}{[\bf $g_{\a_1^2,\a_2^2,\pi}\bu 0$ for $n=1$]}\

 Let $\a_1, \a_2$ be real constants, and $F(x,t,\l)$, $q=0$, and $\pi$ as in Example \ref{hp}.   We apply Theorem \ref{bea} with $g_{\a_1^2,\a_2^2,\pi}$ to the trivial solution $q=0$, $F$ of the third $\hat B_1^{(1)}$-flow.   
 Let 
$$\ti p_1(x,t)= F(x,t,\a_1^2)^{-1}(e_1), \quad \ti p_i(x,t)= F(x,t,\a_2^2)^{-1}(e_i), \, i=2,3,$$
and $\ti\pi(x,t)$ the projection onto $\R \ti p_1(x,t)$ along $\R \ti p_2(x,t) \oplus \R \ti p_3(x,t)$. So 
$\ti \pi = (\ti p_1, 0, 0)(\ti p_1, \ti p_2, \ti p_3)^{-1}$.
After a long computation we obtain a new solution  
$$\ti q_{\a_1, \a_2}:= g_{\a_1^2, \a_2^2, \pi}\bu 0 = (q_1)_{\a_1, \a_2}(e_{11}-e_{33}) +(q_2)_{\a_1, \a_2}(e_{12}+ e_{23})$$  of the third $\hat B_1^{(1)}$ flow, where 
\begin{align*}
& (q_1)_{\a_1, \a_2}=\frac{(\a_1^2-\a_2^2)(\a_2(c_{\a_1}-1)s_{\a_2}-\a_1s_{\a_1}(c_{\a_2}+1))}{4\a_1^2 D_{\a_1. \a_2}}, \\
& (q_2)_{\a_1, \a_2}=\frac{(\a_1^2-\a_2^2)(\a_2^2(c_{\a_1}-1)(c_{\a_2}-1)-\a_1^2(c_{\a_1}+1)(c_{\a_2}+1))}{8\a_1^2D_{\a_1. \a_2}},
\end{align*}
\beq\label{hy}
D_{\a_1,\a_2}=\frac{1}{4}(c_{\a_1}+1)(c_{\a_2}+1)-\frac{\a_2}{2 \a_1}s_{\a_1}s_{\a_2}+\frac{\a_2^2}{4 \a_1^2}(c_{\a_1}-1)(c_{\a_2}-1),
\eeq
and $c_a, s_\a$ are functions defined by \eqref{cv}.  

Note that 
\begin{align*}
c_{\a_i}(x,t)+1& =2\cosh^2(\frac{1}{2}(\a_i x + \a_i^3 t)) \\
c_{\a_i}(x,t)-1 & =2\sinh^2(\frac{1}{2}(\a_i x + \a_i^3 t)) \\
s_{\a_i}(x,t) & = 2\cosh(\frac{1}{2}(\a_i x + \a_i^3 t))\sinh(\frac{1}{2}(\a_i x + \a_i^3 t)).
\end{align*}
 Use Corollary \ref{at} and a direct computation to see that 
\begin{align}
y_{\a_1,\a_2}& =(q_2)_{\a_1, \a_2}+\frac{1}{2}(q_1)_{\a_1, \a_2}^2+\p_x(q_1)_{\a_1, \a_2} \notag \\
&= -2(\ln\mid D_{\a_1, \a_2}\mid)_{xx}  \notag\\
&= (\a_2^2-\a_1^2)\frac{\a_1^2\cosh^2(A_2)+\a_2^2\sinh^2(A_1)}{(\a_1\cosh(A_1)\cosh(A_2)-\a_2\sinh(A_1)\sinh(A_2))^2} \label{bla}
\end{align}
is a solution of the KdV, where $A_i=\frac{1}{2}(\a_i x + \a_i^3 t)$ for $i=1,2$.  Note that we have the following:
\ben
\item When $\a_1 > \a_2 $,  $D_{\a_1, \a_2}$ never vanishes and $y_{\a_1, \a_2}$ is a 2-soliton solution. 
\item When $\a_1 < \a_2 $,  $D_{\a_1, \a_2}$ has zeros and $y_{\a_1, \a_2}$ has singular points. 
\item When $\a_1 = \a_2 $, we get zero solution. 
\item Let $\a_1 \to 0$ and compute directly, we have 
\beq\label{bh2}
\lim_{\a_1\to 0} y_{\a_1, \a_2}=\hat y_{\a_2}=\frac{\a_2^2(4\cosh^2(\frac{1}{2}(\a_2x+\a_2^3t))+\a_2^2x^2)}{(2\cosh(\frac{1}{2}(\a_2x+\a_2^3t))-\a_2x\sinh(\frac{1}{2}(\a_2x+\a_2^3t)))^2}.
\eeq
\item 
Let $\a_2\to 0$ and compute directly to get 
\beq\label{bh}
\lim_{\a_2\to 0} y_{\a_1, \a_2}= \hat y_{\a_1}=-\a_1^2\sech^2(\frac{\a_1}{2}x+\frac{\a_1^3}{2}t),
\eeq
which is a 1-soliton solution of the KdV.
\een

 Next we write down solutions of the isotropic curve flow \eqref{ma} whose isotropic curvature are the above soliton solutions of the KdV.  First by \eqref{bda}, we have 
 $g_{\a_1,\a_2,\pi}^{-1}(0) = \I_{2n+1}-\frac{\a_1-\a_2}{\a_1}\pi -\frac{\a_2-\a_1}{\a_2}\pi^\sharp$. 
By Theorem  \ref{kj}, 
$F(x,t,0)e_1= e^{bx} e_1= (1, x, \frac{x^2}{2})^t$
is a solution of the third isotropic curve flow \eqref{ma}  of B-type on $\cm_{2,1}$ with zero isotropic curvature. 

  By Theorem \ref{mw}(1), we see that
\begin{align*}
\ti \g_{\a_1, \a_2}(x,t):&= F(x,t,0)g_{\a_1^2, \a_2^2, \ti\pi(x,t)}(0)^{-1}e_1\\
&= e^{bx} (\I_{2n+1}- \ti\pi(x,t)-\ti\pi^\sharp(x,t) + \frac{\a_1^2}{\a_2^2} \ti\pi^\sharp(x,t) +\frac{\a_2^2}{\a_1^2} \ti \pi(x,t))
\end{align*}
is a new solution of \eqref{ma} with $y_{\a_1,\a_2}$ as its isotropic curvature. Use the formula for $\ti\pi$ and a direct computation to see that
$$\ti \g_{\a_1, \a_2}=\frac{\a_1^2-\a_2^2}{2 \a_1^2 D_{\a_1, \a_2}} \bpm \frac{\a_1^2}{\a_2^2} (\xi_1)_{\a_1,\a_2} \\
(\xi_1)_{\a_1,\a_2}x +(\xi_2)_{\a_1,\a_2} \\ \frac{\a_2^2}{2\a_1^2}(\xi_1)_{\a_1,\a_2}x^2+ \frac{\a_2^2}{\a_1^2}(\xi_2)_{\a_1,\a_2}x+\frac{\a_2^2}{\a_1^2}(\xi_3)_{\a_1,\a_2} \epm$$
is a solution of \eqref{ma} with isotropic curvature $y_{\a_1, \a_2}$ given by \eqref{bla},
where
\begin{align*}
(\xi_1)_{\a_1,\a_2} & =  \frac{2 \a_1^2 D_{\a_1, \a_2}}{\a_1^2-\a_2^2} - c_{\a_1}-c_{\a_2}, \\
(\xi_2)_{\a_1,\a_2} & = \frac{1}{\a_1}s_{\a_1}(c_{\a_2}+1)-\frac{1}{\a_2}(c_{\a_1}-1)s_{\a_2}, \\
(\xi_3)_{\a_1,\a_2} & = \frac{\a_1^2-\a_2^2}{\a_1^2\a_2^2}(c_{\a_1}-1)(c_{\a_2}+1).
\end{align*}  
(Recall that $c_\a$ and $s_\a$ are defined by \eqref{cv} and $D_{\a_1, \a_2}$ is given in \eqref{hy}). 
\eeg

\beg{[\bf $g_{\a^4, 0, \pi}\bu 0$ for $n=2$]}\ 

In this case, solutions of the third $\hat B_{2}^{(1)}-$ and $\hat B_{2}^{(1)}$ -KdV flow are of the following forms respectively,
\begin{align*}
&q=q_1(e_{11}-e_{55})+q_2(e_{22}-e_{44})+q_3(e_{23}+e_{34}) \\
 & \quad + q_4(e_{13}-e_{35})+q_5(e_{14}+e_{25}),\\
& u= u_1(e_{23}+e_{34})+u_2(e_{14}+e_{45}),
\end{align*}
where by Theorem \ref{mb}, $u$ and $q$ are related by the following formula:
\begin{align}
& u_1= q_3+2(q_1)_x+(q_2)_x+\frac{1}{2}(q_1^2+q_2^2),  \notag \\ 
& u_2= q_5+(q_4)_x+(q_1)_x^{(3)}+q_4(q_1+q_2)-q_3(q_1^2+2(q_1)_x) \notag \\ 
& \quad -(q_1)_xq_2^2-\frac{1}{2}q_1(q_2^2)_x +q_1(q_1-q_2)_x^{(2)}-q_1^2(q_2)_x \notag \\
  & \quad -\frac{1}{2}((q_1)_x)^2-\frac{1}{2}q_1^2q_2^2-2(q_1)_x(q_2)_x-q_1(q_3)_x. \label{bha}
\end{align}

The frame of solution $q=0$ of the third $\hat B_2^{(1)}$-flow with $F(0, 0, \l)=\I_5$ is 
{\footnotesize
\begin{align*}
& F(x, t, \l)  =\exp (J_B(\l)x+J_B^3(\l)t)=F(x, t, z^4) \\
 & =\frac{1}{4} \bpm 2+c_z+C_z  & z(s_z-S_z) & z^2(c_z-C_z) & z^3(s_z+S_z)  & 2(c_z+C_z)-1 \\ \frac{2}{z}(s_z+S_z) & 2(c_z+C_z) & 2z(s_z-S_z) & 2z^2(c_z-C_z) & z^3(s_z+S_z) \\
\frac{2}{z^2}(c_z-C_z) & \frac{2}{z}(s_z+S_z) & 2(c_z+C_z) & 2z(s_z-S_z) & z^2(c_z-C_z) \\
\frac{2}{z^3}(s_z-S_z) & \frac{2}{z^2}(c_z-C_z) & \frac{2}{z}(s_z+S_z) & 2(c_z+C_z) & z(s_z-S_z)  \\
\frac{2}{z^4}(c_z+C_z-2) & \frac{2}{z^3}(s_z-S_z) & \frac{2}{z^2}(c_z-C_z) & \frac{2}{z}(s_z+S_z) & 2+c_z+C_z\epm,
\end{align*}}
where  $\l=z^4$ and 
\begin{align}
& c_z(x, t)=\cosh(z x+ z^3t), \quad   s_z(x, t)=\sinh(z x + z^3 t), \label{fa}\\
& C_z(x, t)=\cos(z x - z^3t), \quad  S_z(x, t)=\sin(z x - z^3 t). \label{fb}
\end{align}

Let $\{ e_1, e_2, \ldots, e_5\}$ be the standard basis of $\R^5$, and $\pi$ the projection onto $\R e_1$ along $\oplus_{i=2}^5 e_i$. Let $\a$ be the nonzero real constant. We apply Theorem \ref{bea} to $q=0$ with frame $F(x, t, \a^4)$ and $g_{\a^4, 0, \pi}$. 

Let 
\begin{equation*}
\ti p_1(x, t)=F(x, t, \a^4)^{-1}e_1, \quad  \ti p_i (x, t)=F(x, t, 0)^{-1}e_i, i=2, \ldots, 5,
\end{equation*}
and $\ti \pi(x, t)$ the projection onto $\R \ti p_1(x, t)$ along $\oplus_{i=2}^5\R p_i(x, t)$.  

 Hence
 {\footnotesize 
\begin{align*} 
\ti p_1=\frac{1}{4}\bpm 2+c_\a+C_\a \\ -\frac{2}{\a}(s_\a+S_\a) \\ \frac{2}{\a^2}(c_\a-C_\a) \\
-\frac{2}{\a^3}(s_\a-S_\a) \\
\frac{2}{\a^4}(c_\a+C_\a-2)
\epm,  \ti p_2=\bpm 0 \\ 1 \\ -x \\ \frac{1}{2}x^2 \\ -\frac{1}{6}x^3-t \epm, \ti p_3=\bpm 0 \\ 0 \\ 1 \\ -x \\ \frac{1}{2}x^2 \epm, \ti p_4=\bpm 0 \\ 0 \\ 0 \\ 1 \\ -x \epm, \ti p_5=e_5.
\end{align*}
}
By Theorem \ref{bea}, we get a new solution $\ti q_\a:= g_{\a^4, 0, \pi} \bu 0$ with
\beq\label{fc}
\bca (\ti q_1)_\a=(\ti q_2)_\a=\a \frac{S_\a-s_\a}{2+c_\a+C_a},\\
 (\ti q_3)_\a=\a^2\frac{C_\a-c_\a}{2+c_\a+C_\a}, \\ (\ti q_4)_\a=0,\\ (\ti q_5)_\a=-\frac{\a^4}{2}.
 \eca
\eeq
Use \eqref{bha} and a long direct computation, we see that 
\begin{align*}
\ti u_1=8\a^2\frac{C_\a-c_\a-S_\a s_\a}{(2+c_\a+C_\a)^2}, \quad \ti u_2=-8\a^4\frac{1}{(2+c_\a+C_\a)^2}.
\end{align*}
 is a 1-soliton solution of the third $\hat B_{1}^{(2)}$-KdV flow \eqref{ef} for any constant $\a\in \R\bh 0$, where $c_\a, s_\a, C_\a, S_\a$ are defined by \eqref{fa} and \eqref{fb}.
 
It follows from Theorem \ref{mw} and a straight forward but long computation that we obtain the curve flow solution of \eqref{jb} whose isotropic curvatures are the 1-soliton solutions $\ti u_1$ and $\ti u_2$ obtained above: 
 \begin{align*}
 \ti \g_{\a^4, 0}(x, t)=\bpm1+x\ti q_1-t\ti q_3+\frac{\a^4}{4}t^2B \\
  A-xB \\
  -xA+\frac{2}{\a^4}\ti q_3+(x^2-t)B \\
  \frac{x^2}{2}A-\frac{2}{\a^4}(x\ti q_3+\ti q_1)-(\frac{1}{2}x^3+tx)B \\
  \frac{2}{\a^4}B \epm,
 \end{align*}
 where $\ti q_1$ and $\ti q_3$ are as in \eqref{fc} and 
 $$A=\frac{2(s_\a+S_\a)}{\a(2+c_\a+C_\a)}, \quad B=\frac{c_\a+C_\a-2}{2+c_\a+C_\a}.$$
 
 We like to mention that Darboux transform of the third $\hat B_1^{(2)}$-KdV flow is also constructed in \cite{VM},  and in \cite{XLW} as a reduction case of a general system (but 1-solitons were not obtained).

\eeg
  
\beg\label{ey}{\bf [Explicit soliton solutions for general $n$]}\ 

 The frame for the trivial solution $q=0$ of the $(2j-1)$-th $\bn1$-flow \eqref{ah} with $F(0,0,\l)=\I_{2n+1}$ is 
$$F(x,t,\l)= \exp(J_B(\l) x + J_B^{2j-1}(\l) t).$$ 
Let $\pi_i$ be rank one projections of $\R^{n+1,n}$ for $1\leq i\leq k$, and $\{\a_i, \b_i\n 1\leq i\leq k\}$ distinct real numbers. 
We apply Theorem \ref{bea} (DT) to the trivial solution $q=0$ to get $k$ $1$-soliton solutions 
$$q_i= g_{\a_i,\b_i,\pi_i} \bu 0= (\a_i-\b_i) [\b, \ti\pi_i-\ti\pi_i^\sharp]$$ of \eqref{ah} and 
$$F_i= g_{\a_i,\b_i, \pi_i} Fg_{\a_i,\b_i,\ti\pi_i}^{-1}$$ is the frame of $q_i$ for $1\leq i\leq k$, where 
$\ti\pi_i(x,t)$ is the projection such that 
$$\Im\ti\pi_i(x,t)= F(x,t,\a_i)^{-1}(\Im\pi_i), \quad \Ker\ti\pi_i(x,t)= F(x,t,\b_i)^{-1}(\Ker \pi_i).$$ 
Apply Theorem \ref{ev} to $q_1$, $q_2$ to get 2-soliton solution $q_{12}$ and its frame $F_{12}$. Apply Theorem \ref{ev} to $q_2$, $q_{21}$ and $q_{23}$ to get the 3-soliton solution $q_{123}$ and its frame $F_{123}$. Continue this way to get explicit formulas for $k$-soliton solutions of \eqref{ah} and their frames. 
We apply Corollary \ref{cga} to soliton solutions of the $(2j-1)$-th $\bn1$-flow to obtain soliton solutions of the $(2j-1)$-th $\bn1$-KdV flow \eqref{ai}.

To write down the corresponding k-soliton solutions, we use Theorem \ref{mw}(1) to see that 
\begin{align*}
&\g_i(x,t)= g(x,t)g^{-1}_i(x,t)e_1\\
&\g_{12}(x,t)= g(x,t) g_1^{-1}(x,t)\ti g^{-1}_{12}(x,t)e_1,
\end{align*}
are solutions of the isotropic curve flow \eqref{cb} of B-type, where 
\begin{align*}
& g(x,t)= \exp(bx+ b^{2j-1}t),\\
&g_i(x,t)= g_{\a_i, \b_i, \ti\pi_i^\sharp(x,t)}(0),\\
& g_{12}(x,t)= g_{\a_2, \b_2, \ti\tau^\sharp_2(x,t)}(0),
\end{align*}
 $\pi^\sharp= \rho_n \pi^t \rho_n$, and $\ti\tau_2$ is the projection with 
 $$\Im \ti\tau_2= g_{\a_1,\b_1,\ti\pi_1}(\a_2)(\Im \ti\pi_2), \quad \Ker\ti \tau_2=  g_{\a_1,\b_1,\ti\pi_1}(\b_2)(\Ker \ti\pi_2).$$ The isotropic curvature of $\g_i$ and $\g_{12}$  are $1$ soliton $q_i$ and $2$ soliton solution $q_{12}$ are given above. 
\eeg

\bs

\end{document}